# Multi-Axis Support-Free Printing of Freeform Parts with Lattice Infill Structures

Yamin Li, Kai Tang*, Dong He, Xiangyu Wang

Department of Mechanical and Aerospace Engineering, Hong Kong University of Science and Technology, Clear Water Bay, Kowloon, Hong Kong

*Corresponding author.

E-mail: ylifm@connect.ust.hk (Yamin Li), mektang@ust.hk (Kai Tang)

**Abstract**: In additive manufacturing, infill structures are commonly used to reduce the weight and cost of a solid part. Currently, most infill structure generation methods are based on the conventional 2.5-axis printing configuration, which, although able to satisfy the self-supporting condition on the infills, suffer from the well-known stair-case effect on the finished surface and the need of extensive support for overhang features. In this paper, based on the emerging continuous multi-axis printing configuration, we present a new lattice infill structure generation algorithm, which is able to achieve both the self-supporting condition for the infills and the support-free requirement at the boundary surface of the part. The algorithm critically relies on the use of three mutually orthogonal geodesic distance fields that are embedded in the tetrahedral mesh of the solid model. The intersection between the iso-geodesic distance surfaces of these three geodesic distance fields naturally forms the desired lattice of infill structure, while the density of the infills can be conveniently controlled by adjusting the iso-values. The lattice infill pattern in each curved slicing layer is trimmed to conform to an Eulerian graph so to generate a continuous printing path, which can effectively reduce the nozzle retractions during the printing process. In addition, to cater to the collision-free requirement and to improve the printing efficiency, we also propose a printing sequence optimization algorithm for determining a collision-free order of printing of the connected lattice infills, which seeks to reduce the air-move length of the nozzle. Ample experiments in both computer simulation and physical printing are performed, and the results give a preliminary confirmation of the advantages of our methodology.

# 1. Introduction

Additive manufacturing (AM) technologies have brought a significant change to manufacturing, which makes it possible to fabricate parts with extremely complex features that are otherwise impossible to be produced by traditional means such as machining. Fused deposition modeling (FDM) is one of the most commonly-used types of AM, owing to its low cost and simplicity, which builds a part layer by layer by extruding a molten filament [1] on the layers. Most FDM systems are of the 2.5-axis printing configuration, namely, the geometric model of the part is first sliced into a series of parallel planar layers and then the material is deposited layer by layer along a fixed direction (Z+). The biggest limitation of this configuration is that support structure is usually needed in order to prevent the collapse of material when printing overhang features, which not only causes the extra cost of the printing material and printing time but also subjects the part surface to potential damage when the support is being removed eventually. Moreover, due to the nature of parallel slicing, parts built under 2.5-axis configuration are inherently susceptible to the staircase effect and the consequent poor surface quality.

Infill structures are commonly used in AM to reduce the weight and cost of a solid part. One critical requirement for the design of infill structures is that they must be self-supporting, as it would be extremely difficult or even impossible to remove any interior support after the part is printed. Dong et al. [2] and Tamburrino et al. [3] reviewed the properties and the modeling processes of lattice infill structures in additive manufacturing. Wu et al. [4] used the idea of adaptive rhombic grids to generate infill structures to satisfy the manufacturing requirements on both the overhang-angle and wall-thickness, and the generated infill structures exhibit improved properties of both high stiffness and static stability. Similarly, Lee et al. [5] proposed a method for generating support-free elliptic voids by constructing a Voronoi diagram of ellipses, which aims at not only avoiding the need of interior supports but also achieving better mechanical properties than Wu's rhombic infill structures [4]. Kuipers et al. [6] proposed a new self-supporting infill structure called *CrossFill*, which has the advantage that the extrusion printing paths are continuous and free of self-overlap. Wang et al. [7] presented a support-free hollowing algorithm based on the offsetting operation which can reduce more volume of material than Wu's method [4]. Yang et al. [8] also proposed a hollow-to-fill algorithm based on the voxel model to guarantee the support-free property of inner surfaces for shape optimization. Similar voxel model-based hollowing methods can also be found in [9–12]. Gupta and Krishnamoorthy [13] developed a framework that can guarantee a sparse infill pattern according to a given arbitrary polygonal mesh; in addition, it guarantees the existence of a single, continuous and crossover-free tool path in each layer by using a novel Euler transformation. Additionally, there are also some infill structure generation methods based on topology optimization [14–16] with self-supporting as a constraint. Notwithstanding their richness, all the above infill structure generation methods are based on the conventional 2.5-axis printing configuration, which means that, although the generated infill structures in the interior of the part are self-supporting, exterior supports are still required when printing the surface of overhang features.

To reduce the exterior support, many approaches have been proposed, although they are still limited to the 2.5-axis printing configuration. Hu et al. [17] proposed an orientation-driven shape optimizer which could considerably slim down the support by finding an optimal building direction. Zhou et al. [18] developed a tree-like support structure generation method based on the Lindenmayer system and an Octree. Vanek et al. [19] presented an optimization framework for the reduction of support, which tries to minimize the support material while providing sufficient support. Tricard et al. [20] proposed an interior rib-like support structure to build hollowed parts. Nevertheless, although in certain cases the support volume could be significantly reduced by these methods, they are not able to fully eliminate the need of exterior support, simply due to the nature of 2.5-axis printing configuration.

As a viable solution to the shortcomings of 2.5-axis printing configuration, the 3+2-axis configuration allows a part to be printed with a finite number of different building directions, although for each building direction the printing configuration is still of the 2.5-axis type. Gao et al. [21] proposed a method to properly decompose a model into several sub-parts, which can then be printed consistently with different building directions with a much reduced amount of supporting material. Wei et al. [22] developed a skeleton-based algorithm for partitioning a model into the least number of sub-parts, which can (at least in theory) totally eliminate the need of support. Wu et al. [23] gave a support-effective volume decomposition algorithm that can minimize the surface area of regions with large overhangs. Based on the 3+2-aixs configuration, Bhatt et al. [24] developed an algorithm to print accurate thin-shell parts with no support. Similar research can also be found in [25–31]. Nonetheless, all these improvements are based on the 3+2-axis configuration, which become less effective for freeform parts with complex features and thus lack generality.

The emerging continuous multi-axis printing configuration is perhaps the ultimate solution to the support-free requirement. Basically, on a multi-axis printing platform, we can not only slice the model into non-planar (i.e., curved) layers, but also continuously adjust the nozzle orientation to align with the layer normal, so to restrict the overhang angle below a threshold, thus achieving a total support-free printing (at least in theory). Dai et al. [32] proposed a curved layer decomposition method for multi-axis printing based on the voxel model. This method is considered to be general for printing freeform parts; however, it demands a huge computational cost due to the nature of voxelization. Xu et al. [33] recently presented a curved layer-decomposition algorithm, which first establishes a geodesic distance field on the part surface, then generates a set of closed iso-geodesic contours on the part surface, and finally fills these 3D contours into curved layers. However, one critical problem with their algorithm is that, as the curved layers are obtained by hole-filling the boundary loops, they are easy to intersect with each other when the slicing distance is small. Xu's method uses the geodesic distance field as an intrinsic indicator to slice the part, which shows good generality and robustness. In terms of the calculation of geodesic field, Crane et al. [34] proposed an efficient and robust method for computing geodesics in a Riemannian manifold. Facilitated by a geodesic distance field embedded on the solid model, the part can be decomposed into a set of curved layers [35,36] that might be suitable for multi-axis printing. Additionally, Etienne et al. [37] proposed a curved layer generation method for printing a part on a 3-axis printers, which can effectively eliminate the staircase effect. Their method first uses planar planes to slice a deformed model of the part and then maps the planar layers back to the original model to obtain the corresponding curved layers.

Although the above curved-layer decomposition methods based on continuous multi-axis printing configuration can significantly reduce or even, in most cases, completely eliminate the need of exterior support, they are all for printing the entire solid volume. Actually, to the best of our knowledge, due to the newness of continuous multi-axis printing configuration, so far there has no published reports on how to automatically design an infill structure and generate a printing path for an arbitrary freeform solid such that no support will be required for either the boundary or the infill structure. In this paper, under the continuous multi-axis printing configuration, we present a new methodology for automatically generating an infill structure as well as the accompanying multi-axis printing path for an arbitrary freeform part, which will be support-free for printing both the infill structure and the boundary surface of the part. The outline of the methodology is given next while its details will be presented in the ensuing sections.

First, a new geodesic distance field (GDF) based curved-layer slicing algorithm is proposed. Rather than using only the iso-geodesic contours on the part's surface as the boundary loops of the curved layers as did in [33], we compute the geodesic distance field directly inside the 3D volume of the part, which provides a more natural volume decomposition. Referring to Figure 1, for a three-dimensional manifold (i.e., a watertight a solid), from the base of the part, we can define locally parallel geodesics that will fill the entire manifold, and the *iso-geodesic distance surfaces* (IGDSs) of this 3D field naturally decompose the whole part. Let us call this geodesic distance field the $\gamma$-GDF. Because IGDSs are always perpendicular to the geodesics, the overhang angle at the boundary of any IGDS of $\gamma$-GDF will be significantly reduced, making it possible to print the part without any support. As IGDSs never intersect each other, the tangling issue of potential intersections between close-by curved layers as filled contours is now conveniently averted.

Then, based on $\gamma$-GDF, a new lattice infill structure generation method is proposed, which will enable support-free printing for both the part boundary surface and the generated infills. Assuming that the printing always begins from the base (i.e., the bottom), a series of IGDSs of $\gamma$-GDF, called $\gamma$-IGDSs (as shown in Figure 1 (b)), can be constructed which naturally decompose the entire part. These IGDSs are exactly the sought curved layers on which lattice infills will be planned. To generate the interior lattice infills, two other types of GDF, called $\alpha$-GDF and $\beta$-GDF respectively, are established, which satisfy the desirable orthogonal property – the geodesics of the three fields are mutually orthogonal to each other. Next, the other two clusters of IGDSs – i.e., the $\alpha$-IGDSs and the $\beta$–IGDSs – are generated, and the three clusters of IGDSs (i.e., the $\gamma$-IGDSs, the $\alpha$-IGDSs, and the $\beta$-IGDSs) are orthogonal to each other, as illustrated in Figure 1 (b). For each $\gamma$-IGDS, a lattice infill pattern is formed by the intersection lines (isolines) between this $\gamma$-IGDS and the $\alpha$-IGDSs and the $\beta$-IGDSs. Because the three clusters of IGDSs are orthogonal to each other, the generated lattice pattern on each $\gamma$-IGDS is assured of self-supporting. In the rest of the paper, the symbols "$\gamma$-", "$\alpha$-" and "$\beta$-", or the superscripts $\gamma$, $\alpha$ and $\beta$ will be used to differentiate similar elements regarding the three GDFs.

Finally, to cater to the collision-free requirement and also to improve the printing efficiency, we propose a printing sequence optimization method that aims at effectively reducing the air-move length of the nozzle while upholding the collision-free constraint.

The rest of the paper is organized as follows. In Section 2, the detailed algorithm of curved layer slicing and lattice infill structure generation is presented. Then Section 3 gives the details of the printing sequence optimization method and the printing path planning method for the lattice infill patterns. In Section 4, to validate the proposed methodology, we report the results of both computer simulation and physical printing experiments on several representative freeform parts. Finally, in Section 5, we conclude the paper and offer some discussions.

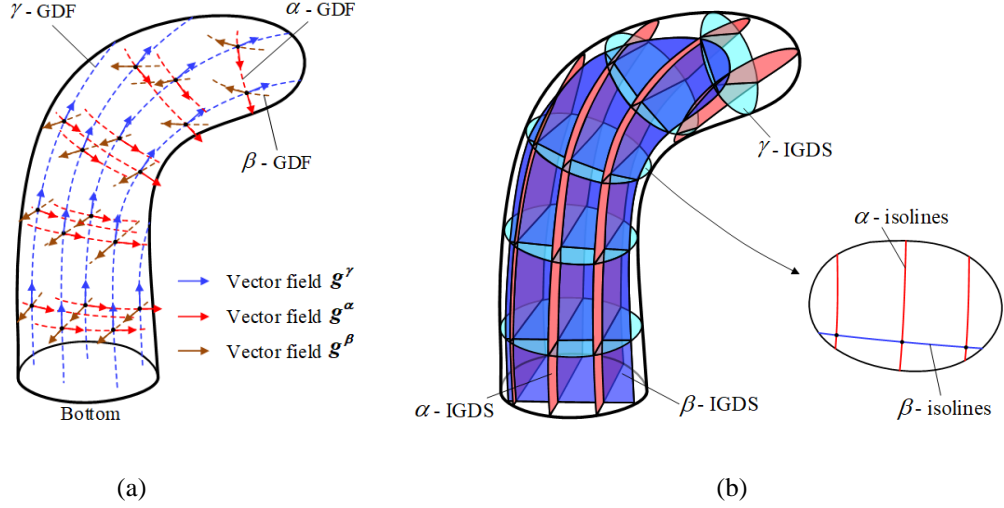

**Figure 1** Illustration of the infill structure generation method: (a) the three geodesic distance fields; (b) the lattice structure formed by the intersection of the three clusters of IGDSs.

## 2. Geodesic distance field-based slicing and infill generation

In this section, we present our GDF-based algorithms for curved layer slicing and generating a self-supporting lattice infill structure. Specifically, in Section 2.1, the details of computing the three mutually orthogonal 3D GDFs on a tetrahedral mesh are given. Then, in Section 2.2, the algorithms for curved layer slicing and the generation of a lattice infill pattern in each layer are presented. Finally, in Section 2.3, the algorithm for the generation of a skeleton tree of the connected lattice infill patterns is proposed, which will facilitate the printing sequence optimization to be presented in Section 3.

### 2.1 Generation of the three mutually orthogonal geodesic distance fields

We assume that the given freeform model is represented by a tetrahedral mesh $M(V, E, F, T)$, where $V$, $E$, $F$ and $T$ are the collections of vertices, edges, faces, and tetrahedrons respectively. First, a 3D GDF embedded on the tetrahedral mesh $M$ is established, where the field value at any vertex is its geodesic distance to the specified bottom of the model, and this GDF is named as $\gamma$-GDF (see Figure 1 (a)). We apply the Crane's heat method on the tetrahedral mesh to calculate $\gamma$-GDF by setting the bottom as the heat source. In their method, the heat flow equation $\dot{u} = \Delta u$ is firstly solved discretely for a fixed time $t$; then, the gradient vector field $X$ can be calculated by $X = -\nabla u/|\nabla u|$; and finally the GDF can be determined by solving the Poisson equation $\Delta \phi = \nabla \cdot X$.

First, to solve the heat diffusion equation $\dot{u} = \Delta u$ on the tetrahedral mesh, it is rewritten in a discrete form as

$$(\mathbf{I} - t\mathbf{V}^{-1}\mathbf{L}_c)\mathbf{u}_t = \mathbf{u}_0 \qquad (1)$$

where $\mathbf{I}$ is the identity matrix, $\boldsymbol{u}_0$ is the initial temperature field vector, $\boldsymbol{u}_t$ is the temperature field vector at moment $t$, $\mathbf{V} \in \mathbb{R}^{n \times n}$ is a diagonal matrix containing the vertex volumes, and $\mathbf{L}_c \in \mathbb{R}^{n \times n}$ is the Laplacian matrix. The detailed values of matrix $\mathbf{L}_c$ can be found in [34]. For the initial temperature field vector $\boldsymbol{u}_0$, the temperatures at the bottom vertices are set as 1, while the temperatures at other vertices are set as 0. The appropriate time step $t$ can be set as $t = h^2$, where $h$ is the average length of edges [34]. Then, the temperature field $\boldsymbol{u}_t$ at moment $t$ can be calculated by solving Eq. (1).

For the $k$th tetrahedron $T_k$, the temperature scalar field inside its volume is defined as a piecewise linear function $u_k(x) = \sum_{i=1}^{4} \varphi_{ik}(x) u_{ik}$, with $\varphi_{ik}$ being the piecewise linear basis function that is valued 1 at vertex $v_i$ and 0 at all other vertices, and $u_{ik}$ being the temperature value at vertex $v_i$. The discrete temperature gradient inside the tetrahedron can then be expressed as

$$\nabla u_k = \sum_{i=1}^{4} u_{ik} \nabla \varphi_{ik} \tag{2}$$

It should be noted that $\nabla \varphi_{ik}$ is simply the vector orthogonal to face $f_{ik}$ and opposite to vertex $v_i$ in tetrahedron $T_k$, pointing towards vertex $v_i$ and with a magnitude of $|\nabla \varphi_{ik}| = \frac{area(f_{ik})}{3|T_k|}$ [38], where $|T_k|$ denotes the volume of tetrahedron $T_k$.

The gradient vector field $\boldsymbol{g}_k^\gamma$ of $\gamma$-GDF can be obtained by normalizing $\nabla u_k$, i.e., $\boldsymbol{g}_k^\gamma = \nabla u_k / |\nabla u_k|$. In order to generate the other two vector fields $\boldsymbol{g}_k^\alpha$ and $\boldsymbol{g}_k^\beta$ which are orthogonal to $\boldsymbol{g}_k^\gamma$, a reference vector $\boldsymbol{r}$ is introduced, which can usually be set as $\boldsymbol{r} = (1 \quad 0 \quad 0)$. Then, $\boldsymbol{g}_k^\alpha$ and $\boldsymbol{g}_k^\beta$ can be calculated by

$$\begin{cases} \boldsymbol{g}_k^\alpha = \frac{\boldsymbol{r} \times \boldsymbol{g}_k^\gamma}{|\boldsymbol{r} \times \boldsymbol{g}_k^\gamma|} \\ \boldsymbol{g}_k^\beta = \boldsymbol{g}_k^\alpha \times \boldsymbol{g}_k^\gamma \end{cases} \tag{3}$$

wherein the vectors $\boldsymbol{g}_k^\gamma$, $\boldsymbol{g}_k^\alpha$ and $\boldsymbol{g}_k^\beta$ are all mutually orthogonal with each other. The integrated divergence of the gradient field associated with vertex $v_i$ can then be written as

$$(Div\boldsymbol{g})(v_i) = \sum_{T_k \in N(i)} \nabla \varphi_{ik} \cdot \boldsymbol{g} |T_k| \tag{4}$$

where $N(i)$ is the set of vertices immediately adjacent to vertex $v_i$. Finally, the three GDFs, i.e., the $\gamma$-GDF $\boldsymbol{\phi}^\gamma$, the $\alpha$-GDF $\boldsymbol{\phi}^\alpha$, and the $\beta$-GDF $\boldsymbol{\phi}^\beta$ for all the vertices can be obtained by solving the following discrete Poisson equation

$$\mathbf{L}_c \boldsymbol{\phi} = \boldsymbol{b} \tag{5}$$

where ***b*** is the divergence field vector of the gradient field.

**2.2 Curved-layer slicing and lattice infill structure generation**

Once the three orthogonal GDFs embedded on the tetrahedral mesh $M$ ($V$, $E$, $F$, $T$) are obtained, the curved layers and the lattice infill pattern in each layer can be constructed. The $\gamma$-GDF is used to slice the whole part, namely, the curved layers are generated by interpolating a number of $\gamma$-IGDSs, and each layer is sandwiched between two adjacent $\gamma$-IGDSs. Let $\Psi^\gamma = \{\phi_1^\gamma, \phi_2^\gamma, \ldots, \phi_i^\gamma, \ldots\}$ be the set of sampling geodesic distances for $\gamma$-GDF. For each $\phi_i^\gamma$, a $\gamma$-IGDS $S_i^\gamma$ can be defined. Similarly, let $\Psi^\alpha = \{\phi_1^\alpha, \phi_2^\alpha, \ldots, \phi_j^\alpha, \ldots\}$ and $\Psi^\beta = \{\phi_1^\beta, \phi_2^\beta, \ldots, \phi_j^\beta, \ldots\}$ be the sets of sampling geodesic distances for the $\alpha$-GDF and $\beta$-GDF respectively, and the corresponding $\alpha$-IGDSs and $\beta$-IGDSs can also be interpolated. The lattice infill pattern in $S_i^\gamma$ can be formed by the intersection lines (i.e., the $\alpha$-isolines) between $S_i^\gamma$ and the $\alpha$-IGDSs, as well as the intersection lines (i.e., the $\beta$-isolines) between $S_i^\gamma$ and the $\beta$-IGDSs. Because the $\gamma$-, $\alpha$-, and $\beta$-IGDSs are all mutually orthogonal to each other, the generated lattice infill pattern in $S_i^\gamma$ is assured of self-supporting if the nozzle orientation is aligned with the normal direction of $S_i^\gamma$.

Figure 2 illustrates an lattice infill pattern $G_i(V_i, E_i)$ at $S_i^\gamma$, which is an undirected graph whose vertex set $V_i$ contains the intersection vertices between the isolines and faces of the tetrahedral mesh, the intersection vertices between the $\alpha$-isolines and the $\beta$-isolines, and the interpolation vertices in terms of $\phi_i^\gamma$ at the mesh boundary, and the edge set $E_i$ is a collection of edges between the vertices.

Algorithm 1 shows the pseudocodes for the generation of $G_i(V_i, E_i)$. Specifically, Steps 1-12 find all the intersection vertices between the mesh faces and the $\alpha$-isolines, wherein a function called $GenGxVertex(F_k, \phi_i^\gamma, \phi_j^\alpha, \&V^\alpha)$ is used to judge whether there exists an intersection vertex $V^\alpha$ between the triangle face $F_k$ and an $\alpha$-isoline in terms of $\phi_j^\alpha$ – it returns true if an intersection vertex is found and false otherwise. Figure 3 illustrates the intersection vertex between an $\alpha$-isoline in terms of $\phi_j^\alpha$ and a face of the tetrahedral mesh. Because the $\alpha$-isolines are on $S_i^\gamma$ which has a $\gamma$-geodesic distance of $\phi_i^\gamma$, there must exist two interpolation vertices in terms of $\phi_i^\gamma$ at the two edges of the triangle face respectively; assuming these two vertices are $P$ and $Q$ at edge $AB$ and $AC$ respectively (see Figure 3), the following condition must be true:

$$\begin{cases} (\phi_a^\gamma - \phi_i^\gamma)(\phi_b^\gamma - \phi_i^\gamma) < 0 \\ (\phi_a^\gamma - \phi_i^\gamma)(\phi_c^\gamma - \phi_i^\gamma) < 0 \end{cases} \quad (6)$$

where $\phi_a^\gamma$, $\phi_b^\gamma$ and $\phi_c^\gamma$ are $\gamma$-geodesic distances at vertex $A$, $B$ and $C$ respectively. Then, the coordinates of vertex $P$ and $Q$, i.e., $v_P$ and $v_Q$, as well as the corresponding $\alpha$-geodesic distances can be calculated by

$$\begin{cases} v_P = (|\phi_a^\gamma - \phi_i^\gamma|v_B + |\phi_b^\gamma - \phi_i^\gamma|v_A)/|\phi_a^\gamma - \phi_b^\gamma| \\ v_Q = (|\phi_a^\gamma - \phi_i^\gamma|v_C + |\phi_c^\gamma - \phi_i^\gamma|v_A)/|\phi_a^\gamma - \phi_c^\gamma| \\ \phi_p^\alpha = (|\phi_a^\gamma - \phi_i^\gamma|\phi_b^\alpha + |\phi_b^\gamma - \phi_i^\gamma|\phi_a^\alpha)/|\phi_a^\gamma - \phi_b^\gamma| \\ \phi_q^\alpha = (|\phi_a^\gamma - \phi_i^\gamma|\phi_c^\alpha + |\phi_c^\gamma - \phi_i^\gamma|\phi_a^\alpha)/|\phi_a^\gamma - \phi_c^\gamma| \end{cases} \quad (7)$$

where $v_A$, $v_B$ and $v_C$ are the coordinates of vertex $A$, $B$ and $C$ respectively, and $\phi_a^\alpha$, $\phi_b^\alpha$ and $\phi_c^\alpha$ are the $\alpha$-geodesic distances at vertex $A$, $B$ and $C$ respectively. The intersection vertex between the triangle face and the $\alpha$-isoline in terms of $\phi_j^\alpha$ must be on edge $PQ$, and the following condition holds:

$$(\phi_p^\alpha - \phi_j^\alpha)(\phi_q^\alpha - \phi_j^\alpha) < 0 \quad (8)$$

Finally, the coordinate $v_N$ of the intersection vertex $N$ can be calculated by

$$v_N = (|\phi_p^\alpha - \phi_j^\alpha|v_Q + |\phi_q^\alpha - \phi_j^\alpha|v_P)/|\phi_p^\alpha - \phi_q^a| \quad (9)$$

Function $GenGxVertex(F_k, \phi_i^\gamma, \phi_j^\alpha, \&V^\alpha)$ first uses Eq. (6) and Eq. (8) to judge whether there exists an intersection vertex at the triangle face $F_k$, and then, if it exists, uses Eq. (7) and Eq. (9) to calculate the coordinate of the intersection vertex $V^\alpha$. After the calculation of all the intersection vertices on the $\alpha$-isoline in terms of $\phi_j^\alpha$, the corresponding edges are constructed by traversing all the tetrahedrons, i.e., if two faces of a tetrahedron contain an $\alpha$-vertex respectively, an $\alpha$-edge is defined to connect these two $\alpha$-vertices (as stipulated in Steps 7-11 in Algorithm 1).

Similarly, Steps 13-24 are used to find all the intersection vertices between the mesh faces and the $\beta$-isolines, wherein the function $GenGyVertex(F_k, \phi_i^\gamma, \phi_j^\beta, \&V^\beta)$ is used to judge whether there exists an intersection vertex $V^\beta$ between the triangle face $F_k$ and the $\beta$-isoline in terms of $\phi_j^\beta$ – it returns true if an intersection vertex is found and false otherwise, and the corresponding $\beta$-edges are constructed by traversing all the tetrahedrons.

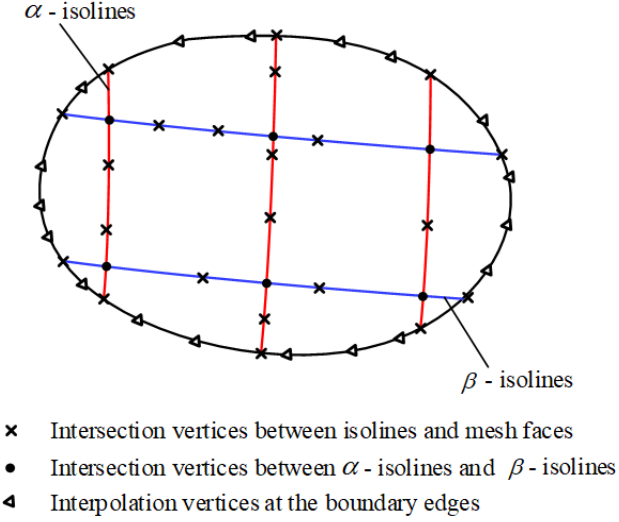

**Figure 2** Illustration of the lattice infill pattern $G_i(N_i, E_i)$ at $S_i^\gamma$

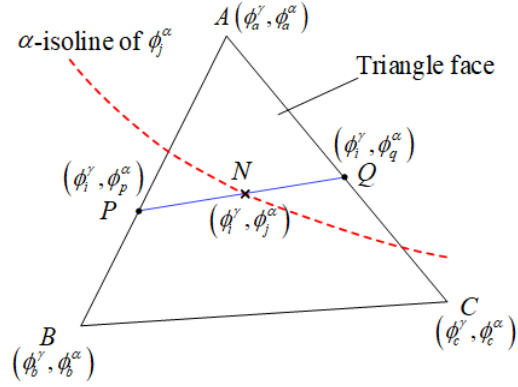

**Figure 3** Illustration of the intersection vertex between an $\alpha$-isoline of $\phi_j^\alpha$ and a face of the tetrahedral mesh

After the generation of all the $\alpha$-edges and the $\beta$-edges, the intersection vertices between them can be obtained by traversing all the tetrahedrons. As shown in Figure 4, a tetrahedron may contain several $\alpha$-edges and $\beta$-edges, and some of them may intersect with each other. Take as an example the intersection vertex $P$ between the $\alpha$-edge $AB$ and the $\beta$-edge $CD$ shown in Figure 4, the following condition must be held:

$$\begin{cases}(\phi_a^\alpha - \phi_{cd}^\alpha)(\phi_b^\alpha - \phi_{cd}^\alpha) < 0 \\ \left(\phi_c^\beta - \phi_{ab}^\beta\right)\left(\phi_d^\beta - \phi_{ab}^\beta\right) < 0\end{cases} \quad (10)$$

where $\phi_a^\alpha$, $\phi_b^\alpha$ and $\phi_{cd}^\alpha$ are the $\alpha$-geodesic distances of node $A$, node $B$, and the $\alpha$-edge $CD$ respectively, while $\phi_c^\beta$, $\phi_d^\beta$ and $\phi_{ab}^\beta$ are the $\beta$-geodesic distances of node $C$, node $D$, and the $\beta$-edge $AB$ respectively. Then the coordinate $v_P$ of the intersection vertex $P$ can be calculated by:

$$v_P = (|\phi_{cd}^\alpha - \phi_a^\alpha|v_B + |\phi_{cd}^\alpha - \phi_b^\alpha|v_A)/|\phi_a^\alpha - \phi_b^\alpha|$$

$$or. \quad v_P = \left(\left|\phi^{\beta}_{ab} - \phi^{\beta}_{d}\right|v_C + \left|\phi^{\beta}_{ab} - \phi^{\beta}_{c}\right|v_D\right) / \left|\phi^{\beta}_{c} - \phi^{\beta}_{d}\right| \tag{11}$$

where $v_A$, $v_B$, $v_C$, and $v_D$ are the coordinates of vertex $A$, $B$, $C$, and $D$, respectively. Inside the tetrahedron $T_k$, all the possible intersection points between the $\alpha$-edges and the $\beta$-edges should be first calculated according to Eq. (10) and Eq. (11). Then, the corresponding edges will be segmented by these intersection vertices, and new edges should be defined according to these intersection vertices. The steps in Algorithm 1 from the 25th row to the 29th are used to find the intersection vertices between the $\alpha$-isolines and the $\beta$-isolines by traversing all the tetrahedrons, where function $GenGxyVertices(T_k, \&VL)$ is used to calculate the intersection vertices $VL$ in the tetrahedron $T_k$ (it returns true if an intersection vertex is found, and false otherwise).

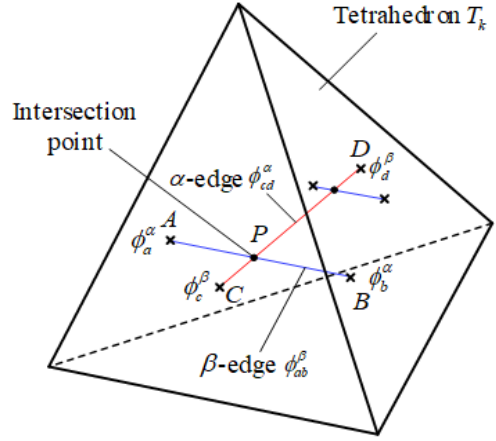

**Figure 4** Illustration of the intersection vertices between the $\alpha$-edges and $\beta$-edges inside a tetrahedron

In Algorithm 1, the steps from the 30th to the 40th row are used to calculate the interpolation vertices in terms of $\phi^{\gamma}_i$ at the boundary of the tetrahedral mesh, as well as to define the corresponding boundary edges for $G_i(N_i, E_i)$. The interpolation vertices can be obtained by traversing all the boundary edges. Specifially, function $GenBoundVertex(E_k, \&V)$ is used to judge whether there is an interpolation vertex $V$ in terms of $\phi^{\gamma}_i$ at the boundary edge $E_k$, which returns true if an interpolation vertex is found, and false otherwise. Take the interpolation vertex $A$ at the boundary edge $MN$ shown in Figure 5 as an example, the following condition must be satisfied:

$$(\phi^{\gamma}_m - \phi^{\gamma}_i)(\phi^{\gamma}_n - \phi^{\gamma}_i) < 0 \tag{12}$$

where $\phi^{\gamma}_m$ and $\phi^{\gamma}_n$ are the $\gamma$-geodesic distances at vertex $M$ and $N$ respectively, and the coordinate $v_A$ of the interpolation vertex $A$ can be calculated by

$$v_A = \left(\left|\phi^{\gamma}_m - \phi^{\gamma}_i\right|v_N + \left|\phi^{\gamma}_n - \phi^{\gamma}_i\right|v_M\right) / \left|\phi^{\alpha}_m - \phi^{a}_n\right| \tag{13}$$

where $v_M$ and $v_N$ are the coordinates of vertex $M$ and $N$ respectively. After the calculation of all the boundary vertices, the corresponding boundary edges can also be established by traversing all the boundary faces. Take the boundary face *MNL* shown in Figure 5 as an example. Vertex $A$ and $B$ are the interpolation vertices in terms of $\phi_i^\gamma$ at edge *MN* and *ML* respectively, with which the edge *AB* can be defined. However, if there are some intersection vertices between face *MNL* and the $\alpha$-edges or the $\beta$-edges, such as point $C$ and $E$ shown in Figure 5, edge *AB* will be broken by these intersection vertices and new edges (i.e., edge *AE*, *EC* and *CB*) should be defined between vertex $A$ and $B$. The above procedure to define the boundary edges in $G_i(N_i, E_i)$ is realized by the steps in the 35$^{th}$ to the 40$^{th}$ row in Algorithm 1, where function *GenBoundEdges*($F_k$, &*EL*) is used to find all the boundary edges at face $F_k$. Finally, all the edges inside the tetrahedrons are saved in the edge list $E_i$ of $G_i(V_i, E_i)$ (i.e., by the steps in the 41$^{st}$ to the 43$^{rd}$ row in Algorithm 1).

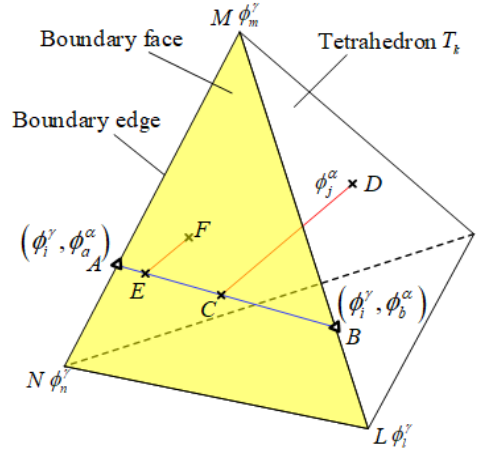

**Figure 5** Illustration of the interpolation points of $\phi_i^\gamma$ at the boundary edges

Algorithm 1  Generation of the lattice infill pattern $G_i$ at $S_i^\gamma$
---
**Input:** the tetrahedral mesh $M(V, E, F, T)$ of the part with three embedded GDFs ($\phi^\gamma$, $\phi^\alpha$ and $\phi^\beta$), $\phi_i^\gamma$, $\Psi^\alpha = \{\phi_1^\alpha, \phi_2^\alpha, \ldots, \phi_i^\alpha, \ldots\}$ and $\Psi^\beta = \{\phi_1^\beta, \phi_2^\beta, \ldots, \phi_i^\beta, \ldots\}$
**Output:** the lattice infill pattern $G_i(V_i, E_i)$ at $\gamma$-IGDS $S_i^\gamma$

```
1    for each ϕⱼᵅ in Ψᵅ
2        for each face Fₖ in F
3            if GenGxVertex(Fₖ, ϕᵢʸ, ϕⱼᵅ, &Vᵅ)
4                Put Vᵅ into Vᵢ, Fₖ ← Vᵅ
5            end
6        end
7        for each tetrahedron Tₖ in T
8            if two faces of Tₖ contain a node V₁ᵅ and V₂ᵅ respectively
9                Build an α-edge Eᵅ of V₁ᵅ and V₂ᵅ, Tₖ ← Eᵅ
10           end
11       end
12   end
```

```
13      for each ϕᵢᵝ in Ψᵝ
14          for each face Fₖ in F
15              if GenGyVertex(Fₖ, ϕᵢᵞ, ϕᵢᵝ, &Vᵝ)
16                  Put Vᵝ into Vᵢ, Fₖ ← Vᵝ
17              end
18          end
19          for each tetrahedron Tₖ in T
20              if two faces of Tₖ contain a nodes V₁ᵝ and V₂ᵝ respectively
21                  Build a β-edge Eᵝ of V₁ᵝ and V₂ᵝ, Tₖ ← Eᵝ
22              end
23          end
24      end
25      for each tetrahedron Tₖ in T
26          if GenGxyVertices(Tₖ, &VL)
27              Put all the vertices in VL into Vᵢ
28          end
29      end
30      for each edge Eₖ in E
31          if GenBoundVertex(Eₖ, &V) .and. Eₖ is a boundary edge
32              Put V into Vᵢ, Eₖ ← V
33          end
34      end
35      for each face Fₖ in F
36          if Fₖ is a boundary face
37              GenBoundEdges(Fₖ, &EL)
38              Put all the edges in EL into Eᵢ
39          end
40      end
41      for each tetrahedron Tₖ in T
42          Put all the edges inside Tₖ into Eᵢ
43      end
```

Take the Y model shown in Figure 6 (a) as an example. The tetrahedral model contains 4014 vertices and 17260 tetrahedrons. The vertices at the bottom are set as the heat source to calculate the $\gamma$-GDF, as well as the $\alpha$-GDF and the $\beta$-GDF. The maximum $\gamma$-geodesic distance is 44.29 mm, and the whole part is sliced into 44 layers with the $\gamma$-geodesic interval set to be 1 mm ($\Psi^\gamma = \{1,2,3...44\}$). At each layer, the lattice infill pattern is generated by Algorithm 1 (the maximum $\alpha$-geodesic distance and $\beta$-geodesic distance are 24.17 mm and 10.11 mm respectively, and the geodesic intervals for these two fields are all set as 2 mm, i.e., $\Psi^\alpha = \{2,4,6...24\}$, $\Psi^\beta = \{2,4,6...10\}$). The generated lattice infill structures are shown in Figure 6 (b-c), and Figure 6 (d) shows the lattice infill pattern at the 24[th] layer. It can be seen that the lattice infill pattern in each layer is self-supporting because both the $\alpha$-GDF and the $\beta$-GDF are orthogonal to the $\gamma$-GDF.

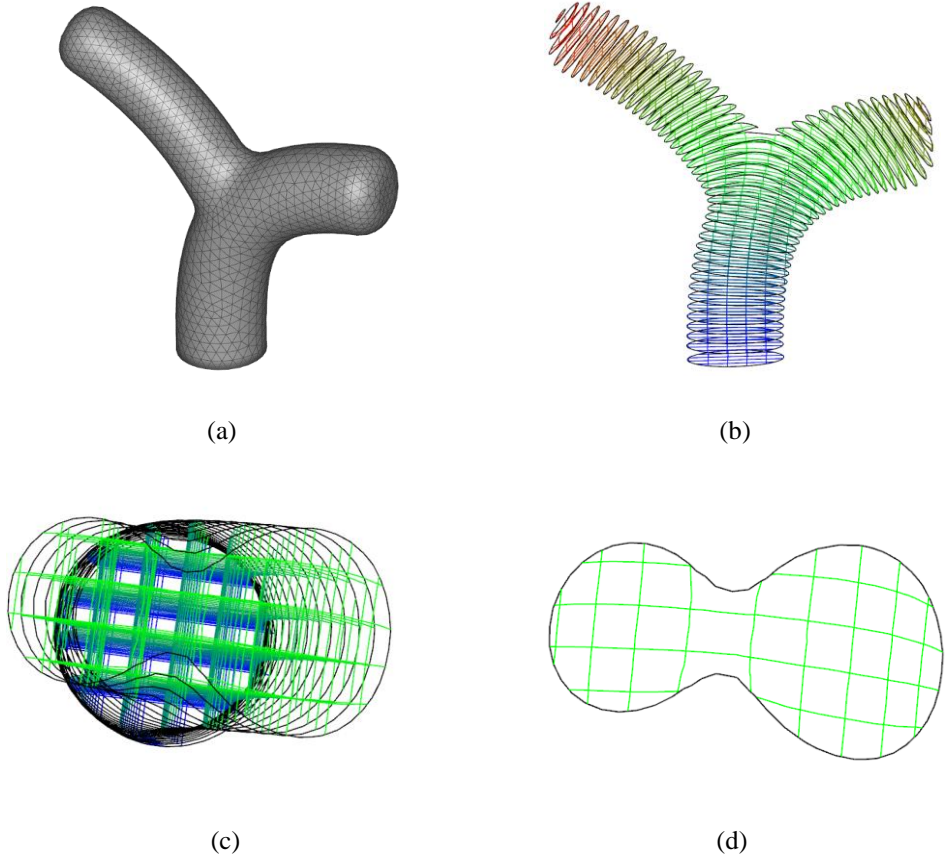

**Figure 6** Generation of lattice infill structures for the Y model: (a) tetrahedral mesh of the model; (b) the generated curved layers; (c) lattice infill structures from the 1st to the 24th layer; (d) lattice infill pattern in the 24th layer.

Referring to Figure 7, the *overhang angle* $\theta$ at the part boundary is defined as the angle between the nozzle orientation and the normal direction $\boldsymbol{n}$ of the boundary surface, which measures the degree of danger of material falling. When angle $\theta$ is larger than a threshold (e.g., 135°), external support structure will be required in order to avoid the falling of material. Figure 8 (a) and (b) show the overhang angles of the Y model under the traditional 2.5-axis planar slicing configuration and our multi-axis GDF based curved-layer slicing method, respectively. As clearly seen, under our method, by aligning the nozzle orientation with the gradient direction of $\gamma$-GDF, angle $\theta$ is kept at or near 90°, thus avoiding the need of extra support.

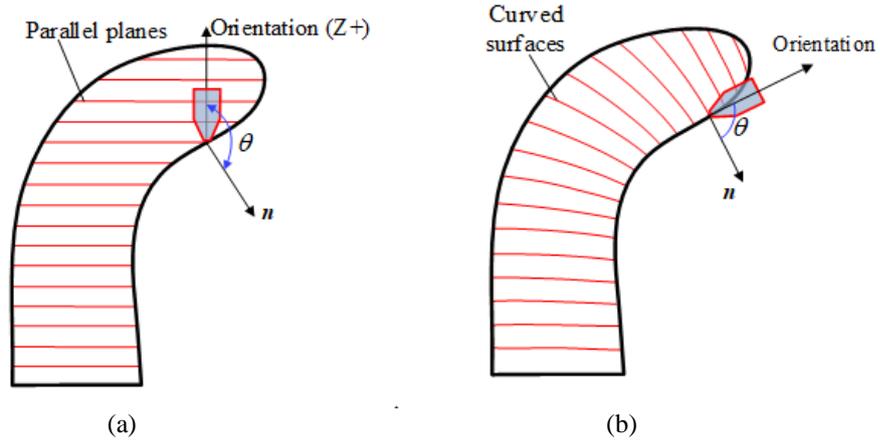

**Figure 7** The overhang angle: (a) traditional 2.5-axis planar slicing; (b) multi-axis curved layer slicing.

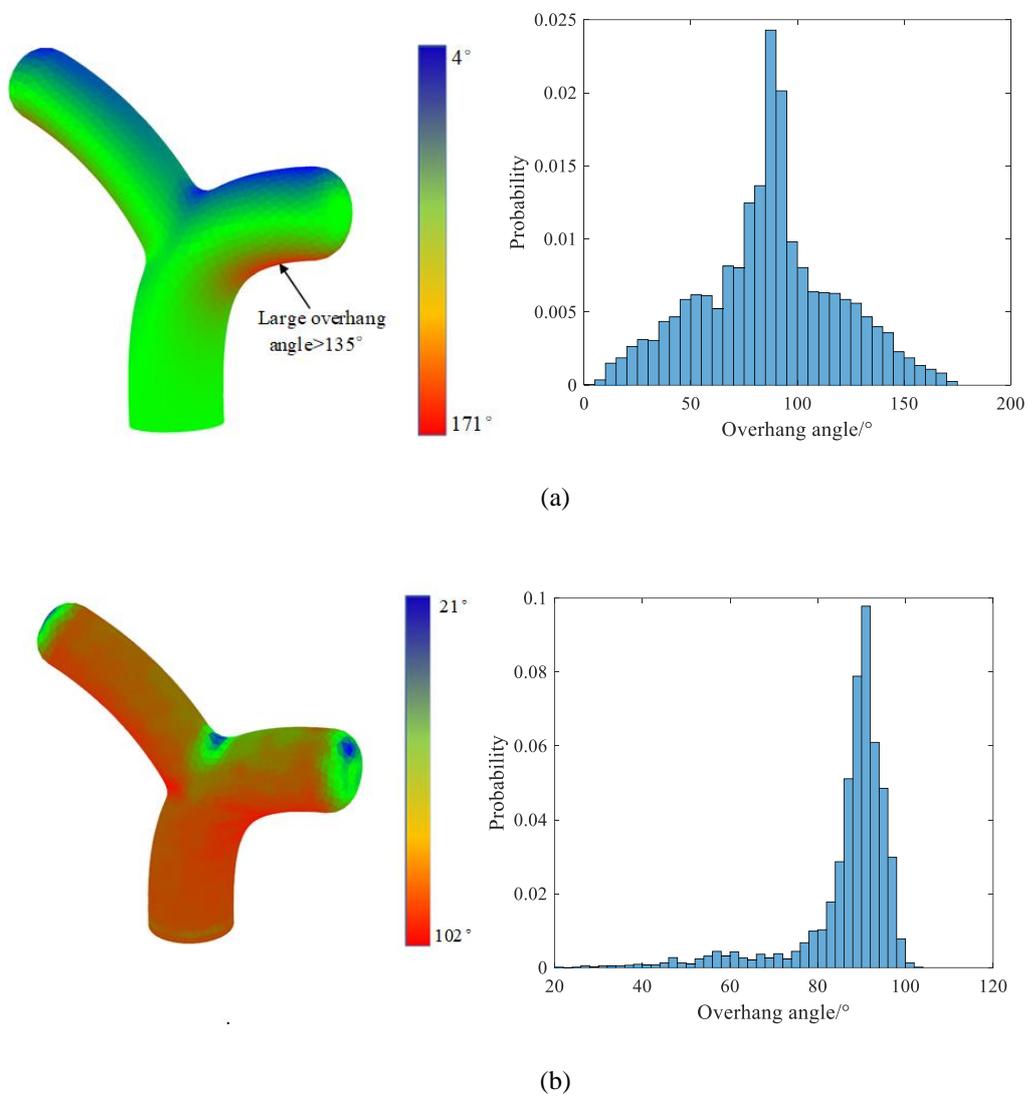

**Figure 8** Overhang angle of the Y model: (a) traditional 2.5-axis slicing method; (b) our multi-axis curved layer slicing method.

## 2.3 The skeleton tree of the connected lattice infill patterns

The generated lattice infill pattern in each layer resembles an undirected graph that may contain several connected sub-graphs, which can be identified by using the DFS (depth first search) algorithm. For all the connected sub-graphs, we define a tree data structure simply called a *skeleton tree* that identifies the topological relationship between them. As illustrated in Figure 9, on the skeleton tree, each node represents a connected sub-graph, and every pair of adjacent sub-graphs are corresponded by an edge between the two representative nodes on the tree. For any node $a$ on the skeleton tree, if node $b$ is connected to $a$ by an edge on the tree, we call $b$ an upper-node of $a$ if $b$'s $\gamma$-geodesic distance is larger than $a$'s, and a lower-node otherwise. For example, on the tree in Figure 9, node $G_{25,1}$ has two upper-nodes, i.e., $G_{26,1}$ and $G_{26,2}$, and only one lower-node, $G_{24,1}$. The skeleton tree is constructed from the bottom towards the top, and the corresponding algorithm is given in Algorithm 1, wherein function *AreTwoGraphsAdjacent* $(G_{i,j}, G_{i+1,k})$ is used to judge whether any two nodes $G_{i,j}$ and $G_{i+1,k}$ are adjacent to each other (i.e., to be connected by an edge). Referring to Figure 10, to judge whether $G_{i,j}$ and $G_{i+1,k}$ are adjacent, we can first randomly select an edge on the triangular mesh (the boundary of the part) that intersects the boundary of $G_{i,j}$; then, from this edge, we trace out a geodesically steepest ascending path on the triangular mesh. Similarly, we can also trace out a geodesically steepest descending path from $G_{i+1,k}$. The two nodes are adjacent to each other if at least one of the two paths go through both.

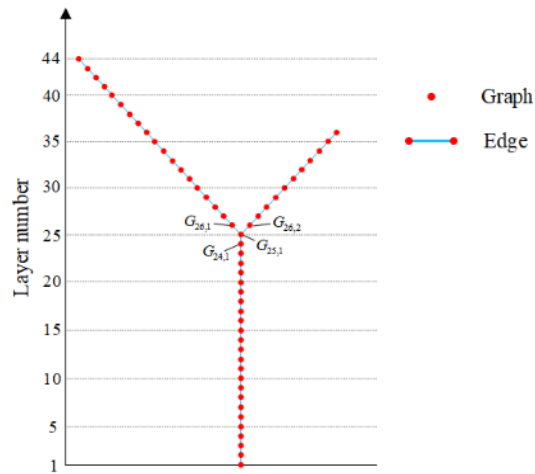

**Figure 9** Generating the skeleton tree of the connected lattice infill patterns

Algorithm 2  Construction of a skeleton tree

**Input:** All the connected sub-graphs of the decomposed layers: $G_i\{G_{i,1}, G_{i,2}, \ldots, G_{i,j}, \ldots\}$, $i = 1, 2, \ldots, n$.
**Output:** Skeleton tree of the connected lattice infill patterns
  1    /* $n$ = the number of layers */

```
2     for i = 1 : n-1
3         for each graph G_{i,j} of ith layer do
4             for each IGDS G_{i+1,k} of (i+1)th layer do
5                 if AreTwoGraphsAdjacent (G_{i,j}, G_{i+1,k}) then
6                     Construct an edge between the two graph nodes
7                 end
8             end
9         end
10    end
```

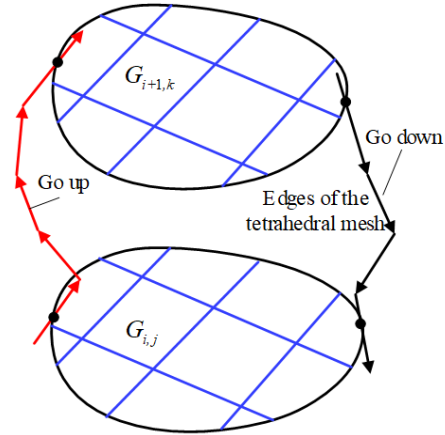

**Figure 10** Judge whether the two sub-graphs are adjacent to each other

## 3. Printing process planning

We have now decomposed the whole part into a number of curved layers according to the $\gamma$-IGDS and constructed the lattice infill pattern in each layer, as well as a skeleton tree that identifies the topological relationship of the connected sub-graphs of the lattice infill patterns. In this section, we will first present our printing sequence generation method (Section 3.1) and then describe how the printing paths are planned (Section 3.2).

### 3.1 Printing sequence generation

The strict increasing geodesic order of the skeleton tree has already defined a partial order of printing – any node must be printed before its upper-node(s). However, since printing is a time-continuous process, we must convert this partial ordering into a total ordering of traversal of the nodes, i.e., a single sequence of nodes to print, called a printing sequence. Hereafter we will interchangeably use the terms a "node" and a connected sub-graph (of a layer). The following criterion must be satisfied for any valid printing sequence:

**Criterion 1**: a sub-graph can only be printed if its lower sub-graph(s) have already been printed.

Generally, there are two traversal strategies of a skeleton tree, i.e. the layer priority traversal (*LPT*) and the depth priority traversal (*DPT*). Refer to Figure 11, the layer priority traversal strategy traverses the skeleton tree layer by layer from bottom-up, which tends to avoid the collision to the utmost, for that the sub-graphs of each layer share the same $\gamma$-geodesic distance. On the other hand, the depth priority traversal strategy traverses the skeleton tree along branches in priority, which favours minimizing the air-movement of the nozzle. However, as shown in Figure 12, under DPT, if a branch grows too deep, it may cause collisions when printing other branches.

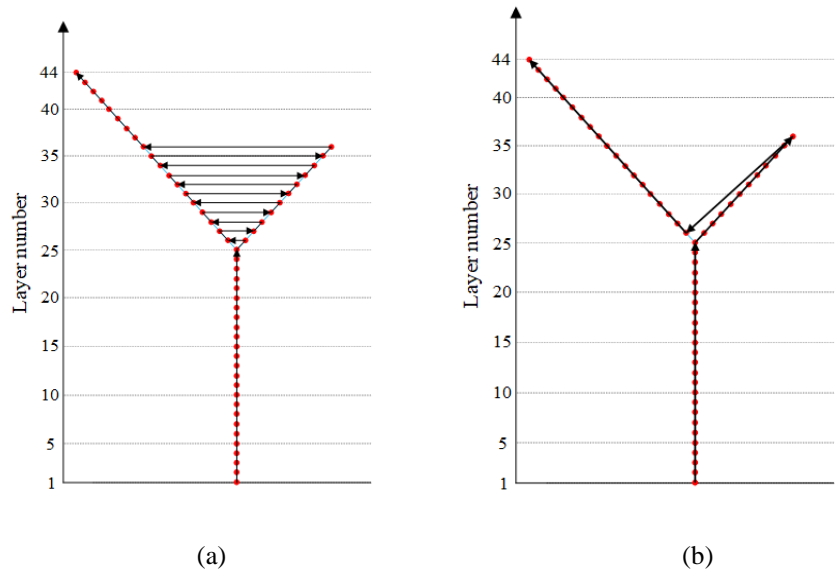

**Figure 11** Traversal strategies of a skeleton tree: (a) layer priority traversal; (b) depth priority traversal.

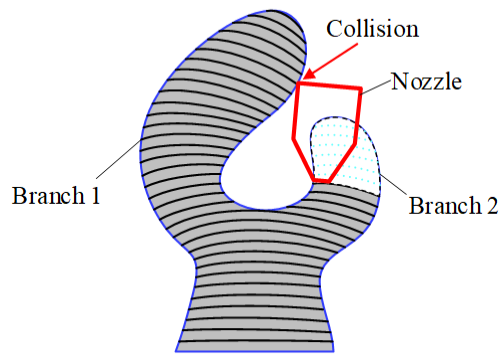

**Figure 12** Illustration of possible collisions during a depth priority traversal

In order to reduce the air-move path length while ensuring no collisions, we propose an optimization strategy which seeks a compromise between LPT and DPT. First of all, the collision check between the nozzle and the sub-graphs must be modelled. In this paper, the shape of the nozzle is simplified by its bounding cone, as shown in Figure 13 (a). Admittedly, this simplification is too conservative; however, because collision check is not the main topic of this paper, we opt for this simplification to implement our algorithm. When the nozzle cone sweeps along the boundary curve of a sub-graph with its orientation coincident with the surface normal direction $n$, the envelope of motion will be a ring-like ruled surface $S(u, v) = P(u) + vk(u)$, where $P(u)$ is an arbitrary point on the boundary curve of the sub-graph, and the unit vector $k(u)$ of the generator can be obtained by rotating the normal vector $n$ at $P(u)$ around the tangent vector $\tau(u)$ of the boundary curve with the nozzle angle $\alpha$, as shown in Figure 13 (a). Specifically, to construct the triangular mesh of the ruled surface, we first place a few sample points on the boundary curve of the sub-graph, then calculate their generators, and finally connect the generators as triangles. The upper and lower holes of the ring-like ruled surface should be filled to approximate the envelope volume of the cone over the entire sub-graph. In this paper, we adopt the advancing front mesh (AFM) technique [39] to fill the holes, which is robust and simple. To determine whether there is a potential collision when printing a sub-graph, we only need to check whether there are intersections between other sub-graphs and this envelope volume. For each sub-graph, we can calculate all the *potential collision sub-graphs* (PCG) (i.e., other sub-graphs that intersect the envelope volume of this sub-graph). Take the part shown in Figure 13 (b) as an example, the potential collision sub-graphs for sub-graph $G_{4,1}$ will be $G_{5,1}$, $G_{6,1}$, $G_{7,1}$, $G_{7,2}$ and $G_{8,2}$. The detailed procedures for calculating the PCGs of each sub-graph are given in Algorithm 3, where function *CollisionCheck* ($G_i$, $G_j$) is used to judge whether there are intersections between sub-graph $G_j$ and the envelope volume of sub-graph $G_i$ – it returns true if an intersection is identified and false otherwise.

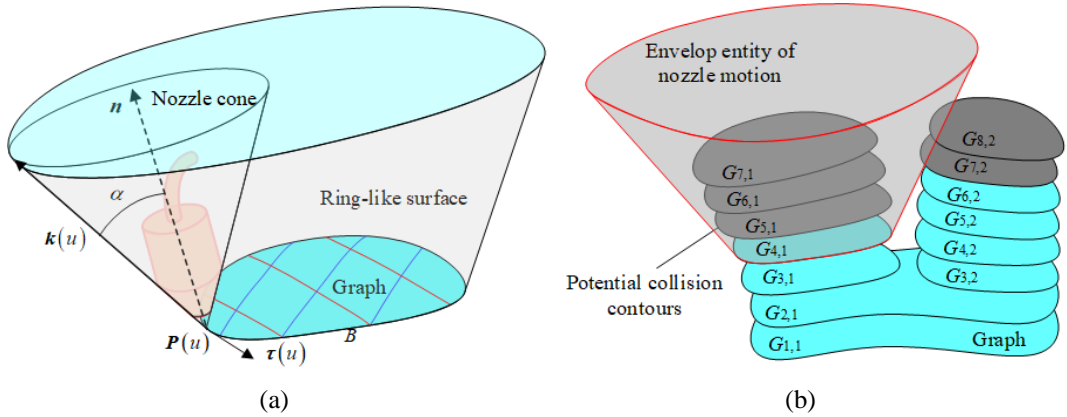

**Figure 13** Illustration of collision check between the nozzle and a sub-graph: (a) ring-like surface generated by sweeping the nozzle along the boundary of the sib-graph; (b) envelope volume of the nozzle motion.

Algorithm 3  Calculation of the potential collision sub-graphs of each sub-graph

**Input:** the graph list $\{G_1, G_2, …, G_i, …\}$ and the nozzle cone angle $\alpha$
**Output:** PCGs list for each graph
1   **integer** $k$ = total number of sub-graphs
2   **for** $i = 1: k$ **do**
3       **for** $j = 1: k$ **do**

```
4        if  i != j then
5            if CollisionCheck (G_i, G_j) then
6                Put G_j into the PCGs list of G_i.
7            end
8        end
9    end
10 end
```

Facilitated by the PCGs of each sub-graph, we propose a greedy traversal (*GT*) algorithm that can generate a collision-free printing sequence with a shorter air-move path length than that of the layer priority traversal. Besides Criterion 1, another criterion must be satisfied during the traversal:

**Criterion 2**: a sub-graph can only be printed if the PCGs of all the unprinted sub-graphs exclude this sub-graph.

Specifically, in Algorithm 4 below, function *UpdateNodeCadidates* (*ST*) is used to find all the printable candidate nodes which satisfy both Criterion 1 and Criterion 2 according to the current skeleton tree *ST*. And function *SelectANode* (*NC*, $N_c$) is used to update the current node $N_c$ from the candidate node list *NC*. It will select the upper-node(s) of $N_c$ in priority. However, if the upper-node(s) are not included in *NC*, the node which is nearest to $N_c$ will be selected. To summarize, we traverse the skeleton tree along the branches in priority unless a potential collision is encountered.

Algorithm 4  Printing sequence optimization algorithm

```
Input: The skeleton tree ST of all the sub-graphs
Output: Printing sequence list PQ of the sub-graphs
1    Node candidates list NC = UpdateNodeCadidates (ST)
2    Current node N_c = either node in NC
3    while  NC != ∅ do
4        Label N_c as printed
5        Put N_c into PQ
6        NC = UpdateNodeCadidates (ST)
7        N_c = SelectANode (NC, N_c)
8    end
```

## 3.2 Printing path planning for lattice infill structures

For any node (sub-graph) in the skeleton tree, a traversal printing path needs to be determined. As illustrated in Figure 14 (a), for a connected sub-graph $G(V,E)$, the number of intersection vertices $(v_1, v_2, \ldots, v_k, \ldots)$ between the boundary curve and the isolines is always even and the degrees of these vertices are all 3, while the degrees of other vertices are either 2 or 4. To avoid excessive tool retractions, the sub-graph can be transformed into an Eulerian graph by properly trimming the boundary curve. As shown in Figure 14 (b), the boundary edges between the intersection vertex $v_k$ and $v_{k+1}$ ($k = 2, 4, 6, \ldots$) are deleted, so that the degrees of all the vertices become even and the new graph must contain an Eulerian tour. In this paper, the well-known Fleury's algorithm is used to find an Eulerian tour in a connected graph. However, to avoid crossovers on the printing path, the printing path will turn at any vertex whenever its degree is 4, i.e., at the intersection vertices between the $\alpha$-isolines and the $\beta$-isolines, as shown in Figure 15. To provide supports at places where the boundary edges are deleted, a support perimeter around the trimmed graph is added, which though is jagged at these places, as shown in Figure 14 (c) and (d). The offset distance between the boundary curve and the support perimeter is set to be the path width $w$, and the tooth length $l$ can be adjusted to a proper value, e.g., $l = 2w$. The nozzle orientation at each vertex is set to be coincident with the layer surface normal vector of the corresponding $\gamma$-IGDS. Because IGDSs are always perpendicular to the geodesics, the gradient vector of the geodesic field at the vertex can be directly used as the nozzle orientation.

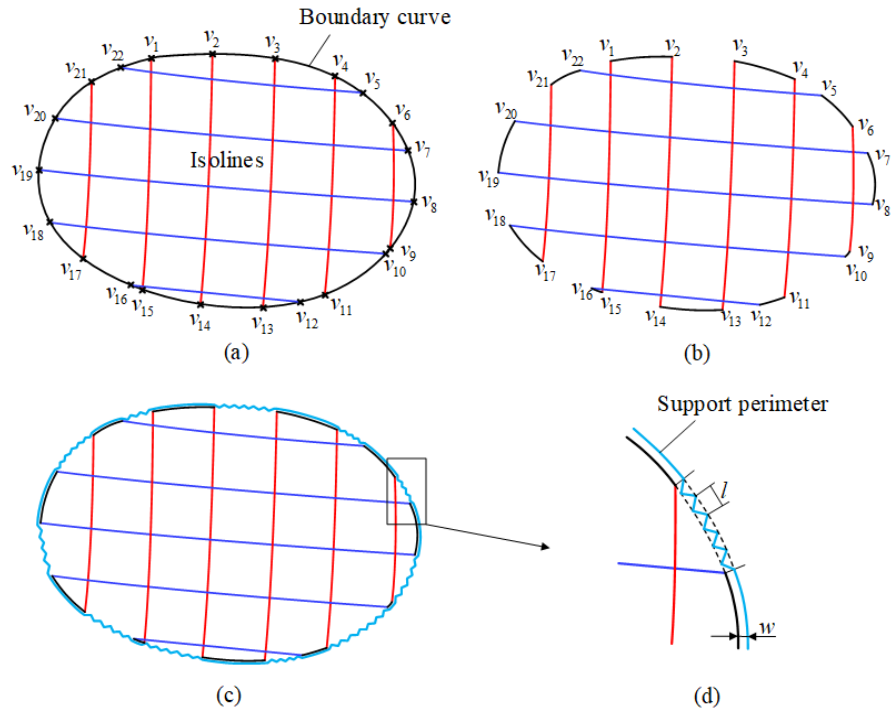

**Figure 14** Printing path for a lattice infill pattern: (a) lattice infill pattern; (b) trimmed pattern that contains an Eulerian tour; (c) printing path for the lattice infill pattern; (d) generation of the support perimeter.

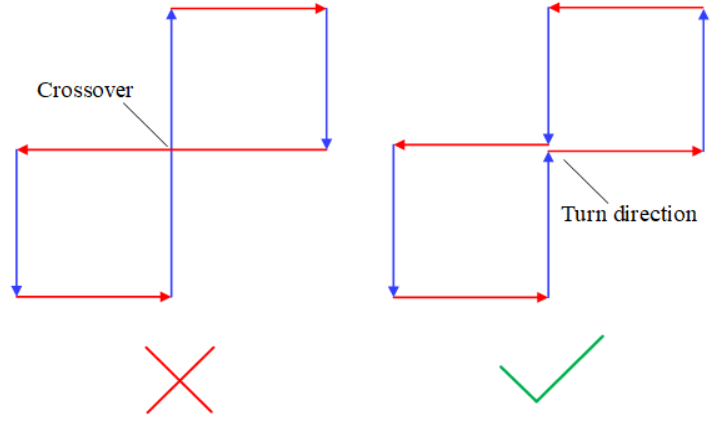

**Figure 15** Avoidance of crossovers

Due to the nature of curved layer slicing, the layer thickness *h* will no longer be a constant along the printing path. Refer to Figure 16, the intersection of the extruded filament is simplified to be a rectangle of *h*×*w*, where *w* is the path width; then, the following mass conservation equation should be satisfied during the printing process:

$$\pi r_m^2 f_m = \mu w h f_p \tag{14}$$

where $r_m$ is the radius of the original filament, $f_m$ is the feed rate of the filament, $f_p$ is the feed rate of the nozzle, and $\mu$ is a correction coefficient which is smaller than 1 and can be determined by experiments. The layer thickness $h_k$ at vertex $v_k$ of the *i*th sub-graph $G_i(V_i, E_i)$ can be calculated by finding the shortest distance of this vertex to the previous (*i*-1)th sub-graph $G_{i-1}(V_{i-1}, E_{i-1})$:

$$\begin{cases} h_k = FindShortestDistance(v_k, G_{i-1}(V_{i-1}, E_{i-1})) \\ if\ h_k \geq \lambda \phi_i, h_k = \lambda \phi_i \end{cases} \tag{15}$$

wherein function $FindShortestDistance(v_k, G_{i-1}(V_{i-1}, E_{i-1}))$ traverses all the edges of $G_{i-1}(V_{i-1}, E_{i-1})$ to find the shortest distance. To avoid a too large layer thickness, $h_k$ should not exceed a threshold $\lambda \phi_i$, where $\phi_i$ is the $\gamma$-geodesic interval between $G_i(V_i, E_i)$ and $G_{i-1}(V_{i-1}, E_{i-1})$, and $\lambda$ is a coefficient which is larger than 1 (1.5 in our tests). Additionally, at the first layer, $h_k$ can be set as the z-coordinate $z_k$ of vertex $v_k$. For the Y model, Figure 17 (a) shows the distribution of layer thickness deviation *e* at different layers when the geodesic distance interval $\phi_i$ is 1mm ($e = (h_k - \phi_i)/\phi_i$), with a maximum percentage deviation of 35%, and Figure 17 (b) shows the statistical results. Although the overhang angle at the boundary of the part is considerably reduced, the layer thickness is nonuniform due to the intrinsic nonuniform distribution of geodesics.

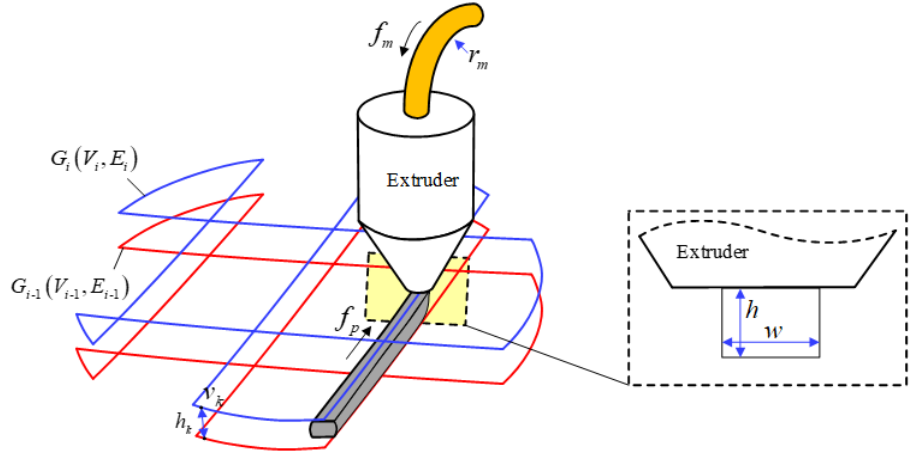

**Figure 16** Illustration of the printing parameters

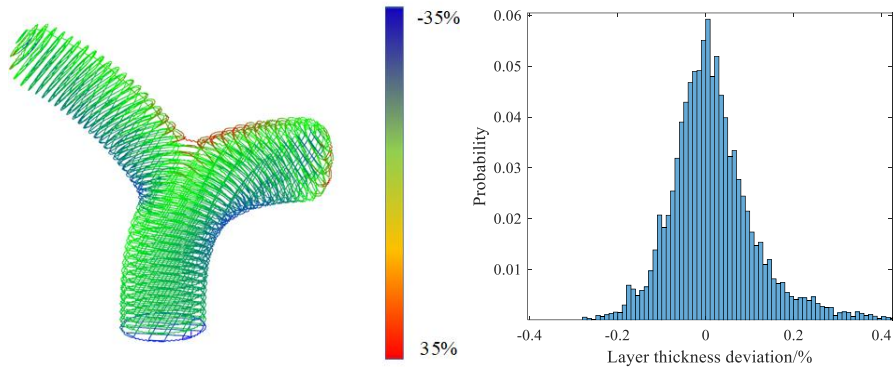

**Figure 17** Illustration of layer thickness of the lattice infill patterns: (a) distribution of layer thickness deviation at different layers; (b) statistical chart of layer thickness deviation at different layers.

To avoid possible collisions when the nozzle moves from one connected sub-graph to another, in this paper, we make use of the safe box method (as reported in our recent works [40,41]). As schematically shown in Figure 18, under the safe box paradigm, the nozzle first air-moves from the current sub-graph to one of the safe planes (which are outside the current in-process workpiece), then moves on this safe plane, crosses the obstacle (i.e., the in-process workpiece), and finally approaches another sub-graph.

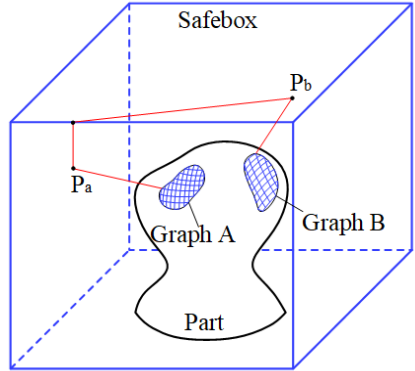

**Figure 18** Safe box of the in-process workpiece and planning of the collision-free air-move of the nozzle

**4. Experiments and discussion**

We have implemented the proposed methodology of automatically generating the lattice infill structures and printing path for multi-axis support-free printing of an arbitrary freeform part in C++ and run the computer program on a laptop with an Intel i7 CPU. In addition, for the purpose of physical validation, as shown in Figure 19, we have built a simple multi-axis FDM printer, which is composed of a 6DOF robot arm (UR5) and a three-axis filament feed system. The robot enables the in-process workpiece to realize any desirable posture with respect to the nozzle, while at the same time the robot and the filament feed rates are synchronously controlled to ensure that Eq. (14) is always satisfied during the printing process. Five freeform parts with complicated structures such as with overhangs and of non-zero genus numbers were chosen for the test and were also physically printed. In this section we report the experimental results of both computer simulation and physical printing, and together with our discussion.

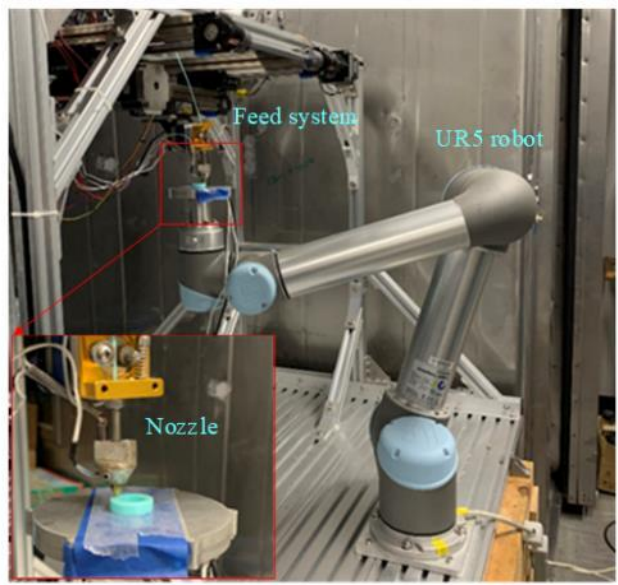

**Figure 19** Homebuilt multi-axis robot printing system

## 4.1 Printing of the freeform parts

Figure 20 depicts the tetrahedral models of the five freeform parts that are printed by using the proposed lattice infilling method, i.e., the Y model, the spiral model, the bunny, the kitten, and the propeller. Table 1 lists the printing parameters of the five parts. The $\gamma$-geodesic distance interval for the curved layer slicing is set to be 0.6 mm. The lattice width for the first four parts are set to be 4 mm, 6 mm, 6mm, and 6 mm respectively, while the propeller is printed in two steps – in the first step, the cylinder is printed with the lattice width set to be 8 mm, and in the second step the three blades are printed with the lattice width set to be 3 mm. Figure 21 and Figure 22 show the actual and simulated printing processes of the five parts, respectively, and Figure 23 shows some cross-sections of three printed parts. All the five freeform parts are successfully printed without any supports, for both the part boundary surface and the interior infills. Due to the intrinsic nature of geodesic distance field, the overhang angle at the part boundary surface is considerably reduced, making it possible to print a part without any exterior supports. As already described, the lattice infill structures are formed by the intersections between the three clusters of orthogonal IGDSs (i.e., the $\gamma$- IGDSs, the $\alpha$-IGDSs and the $\beta$-IGDSs.), which guarantees that the generated infill structures are self-supporting.

In terms of the computational cost, as our volume decomposition method is a combination of several computational processes, Table 2 lists the time complexities of these processes and the actual amounts of running time of the first four tests. The calculation of the geodesic distance field involves solving a linear system, so the time complexity is only $O(n)$, where $n$ is the number of mesh vertices. The time complexity of the lattice infill structures' generation is $O(n*m)$, where $n$ and $m$ are respectively the numbers of sliced layers and mesh vertices, as the time complexity of Algorithm 1 is $O(m)$, and Algorithm 1 will be executed $n$ times to generate the lattice infill for each layer. After the trimming of the lattice infill in each layer, the well-known Fleury's algorithm is used to find an Eulerian tour, whose time complexity is $O(n^2)$, where $n$ is the number of vertices in the sub-graph. The time complexity of layer thickness calculation (Eq. (15)) is $O(n*m)$, where $n$ is the number of vertices of the current sub-graph, and $m$ is the number of edges of the previous sub-graph. By using a *kd*-tree, the time complexity can be reduced to $O(n*log m)$.

Table 1 Printing parameters of the five parts

| Part | Number of tetrahedrons | Time for printing path generation (s) | Path length (mm) | Actual printing time (min) | Number of layers | Width of the lattice (mm) |
|---|---|---|---|---|---|---|
| Y | 17148 | 33 | 49765 | 130 | 145 | 4 |
| Spiral | 31152 | 74 | 71139 | 171 | 401 | 6 |
| Bunny | 66467 | 203 | 194260 | 400 | 206 | 6 |
| Kitten | 58146 | 120 | 88338 | 232 | 161 | 6 |
| Propeller | 83314 | 145 | 125425 | 327 | Cylinder: 183 Blade: 83 | Cylinder: 8 Blade: 3 |

Table 2 Time complexities of the algorithms and the running time of the three tests

| Process | Algorithm | Time complexity | Running time (s) |
|---|---|---|---|
| Geodesic distance field generation | Sec. 2.1 | $O(n)$ | Y: 6; Spiral: 23; Bunny: 48; Kitten: 39 |
| Lattice infill generation | Algorithm 1 | $O(n*m)$ | Y: 5; Spiral: 17; Bunny: 26; Kitten: 16 |
| Finding a Eulerian tour | Fleury's algorithm | $O(n^2)$ | Y: 14; Spiral: 23; Bunny: 77; Kitten: 39 |
| Layer thickness calculation | Eq. (15) | $O(n*m)$ | Y: 4; Spiral: 6; Bunny: 36; Kitten: 15 |

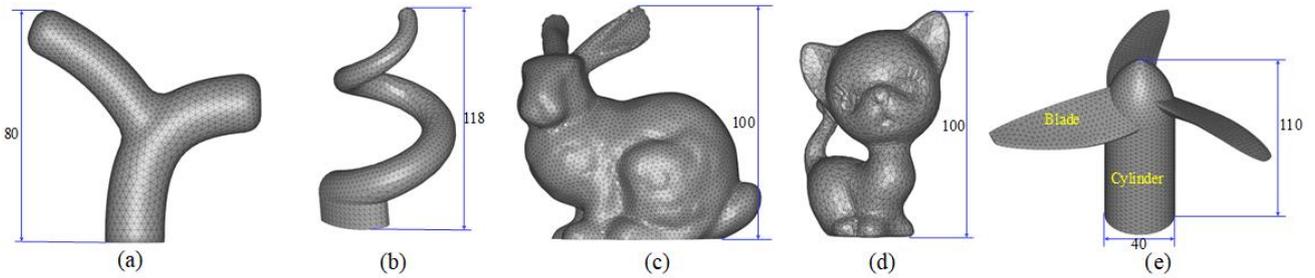

**Figure 20** Five freeform parts to print: (a) Y model; (b) spiral model; (c) bunny; (d) kitten; (e) propeller.

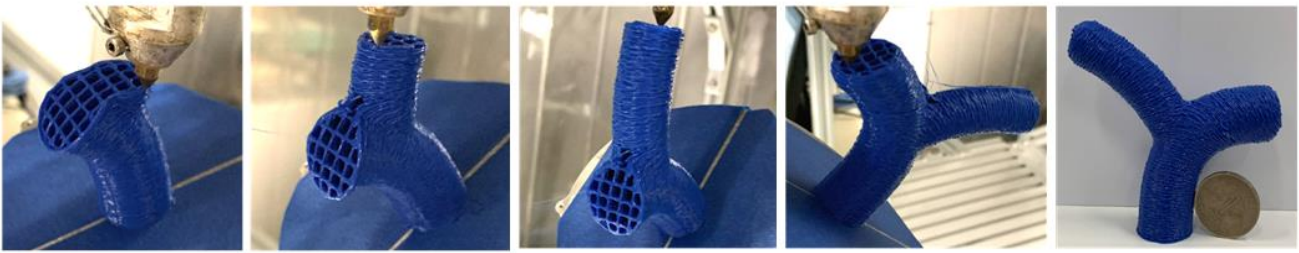

(a)

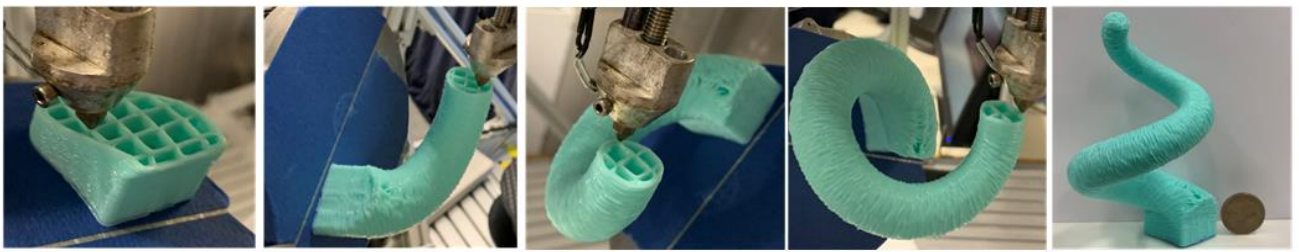

(b)

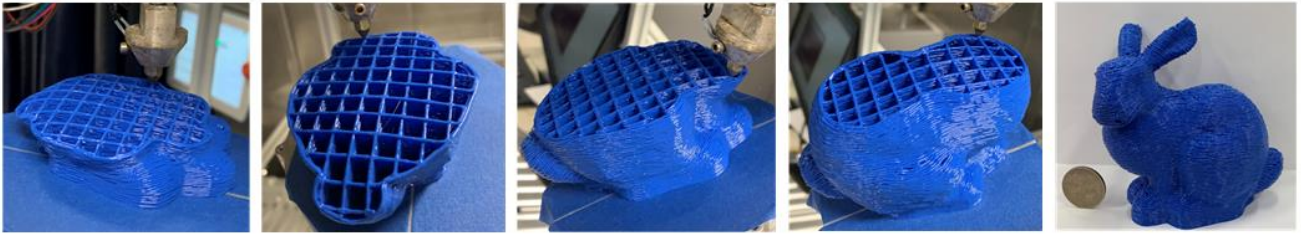

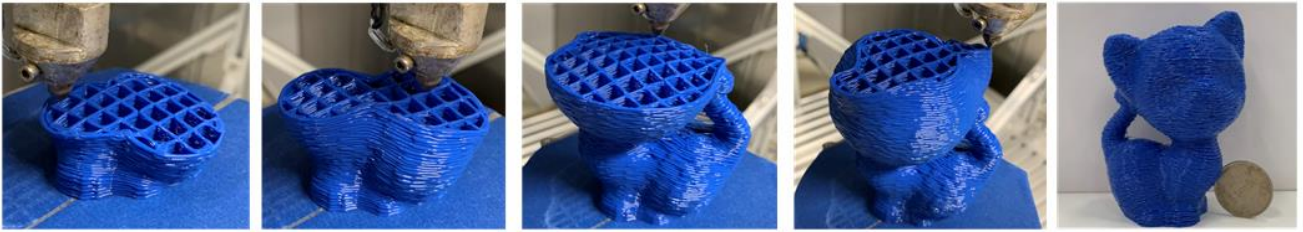

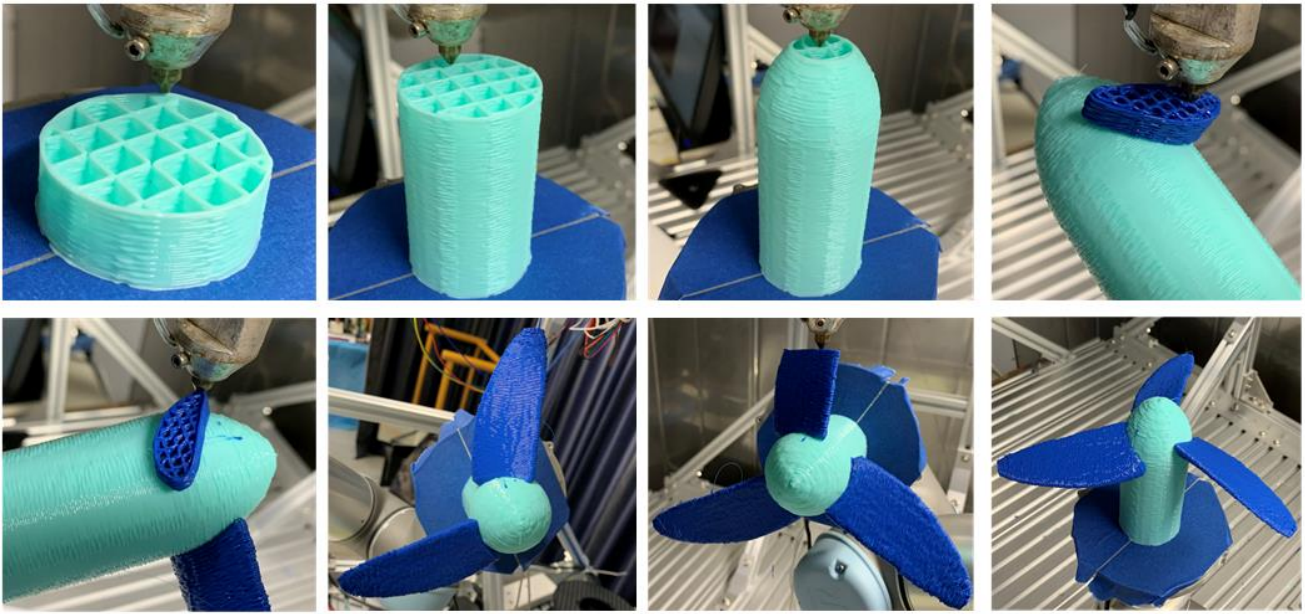

**Figure 21** Multi-axis support-free printing with lattice infill structures: (a) Y model; (b) spiral model; (c) bunny; (d) kitten; (e) propeller.

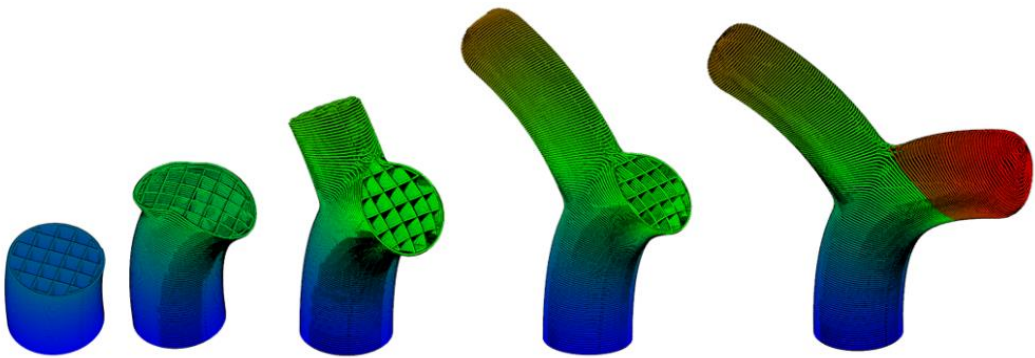

(a)

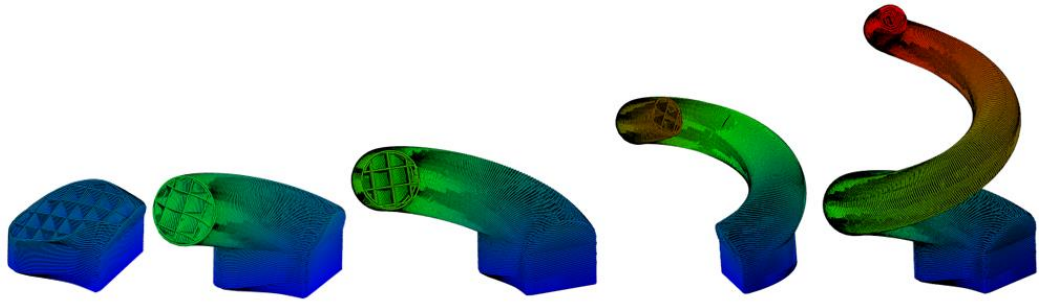

(b)

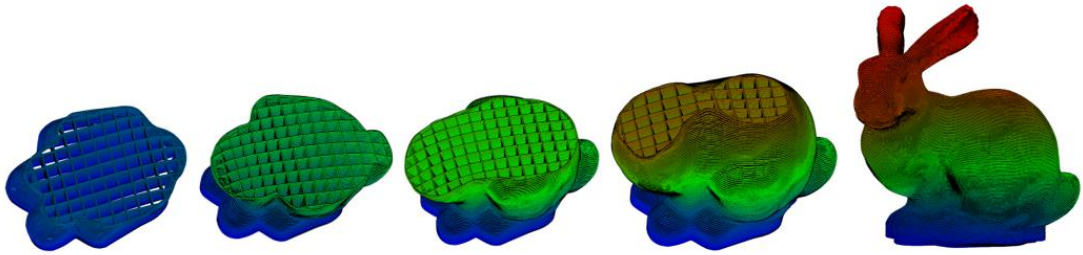

(c)

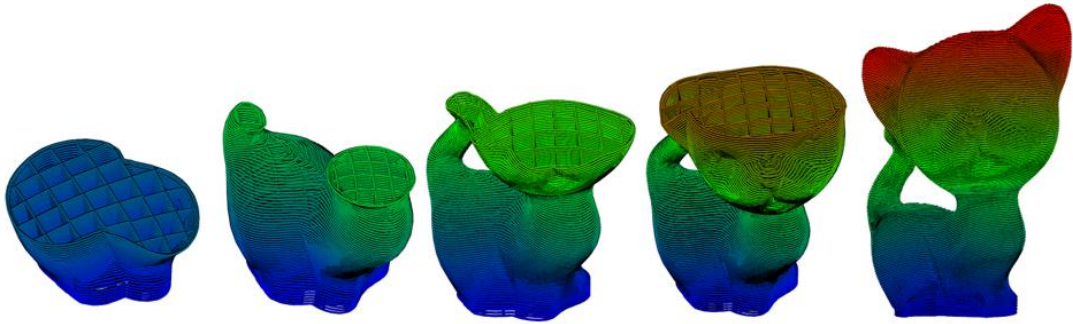

(d)

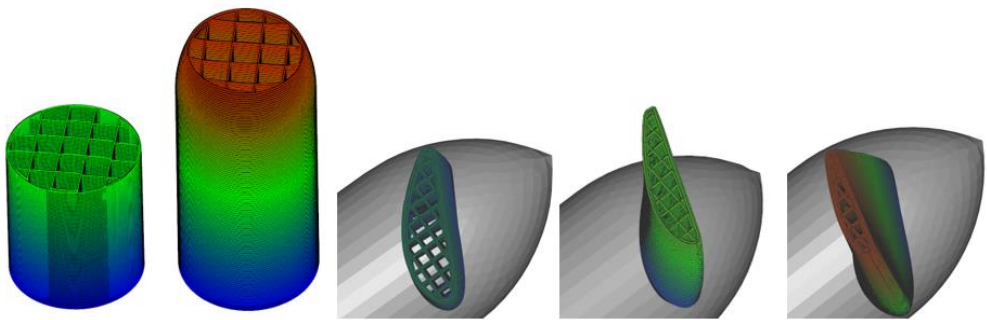

(e)

**Figure 22** Simulation of the printing processes: (a) Y model; (b) spiral model; (c) bunny; (d) kitten; (e) propeller.

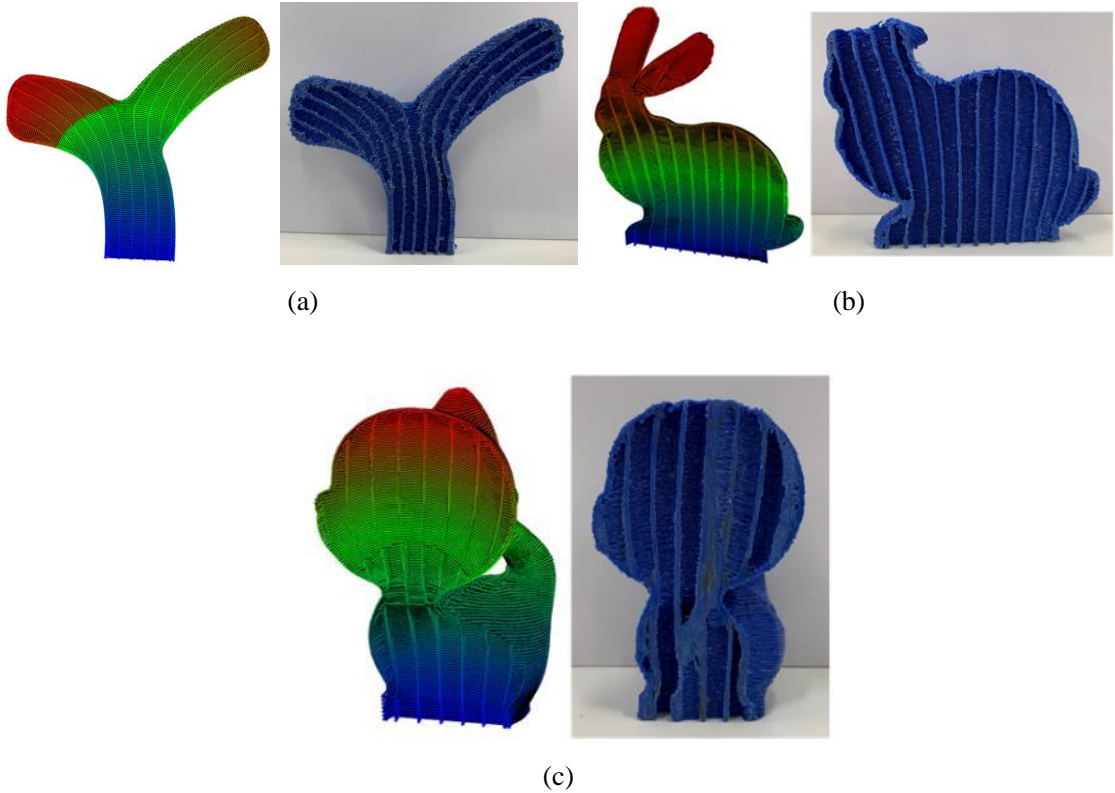

**Figure 23** Cross-sections of the printed parts: (a) Y model; (b) bunny; (c) kitten.

The lattice width (i.e., geodesic intervals for $\alpha$-GDF and $\beta$-GDF) can be set to be a constant, which will result in even lattices (see in Figure 21 and Figure 24 (a)). However, to print a part with graded material properties (e.g., variable Young's modulus), the lattice width can also be adjusted adpatively, namely, by increasing the density of infill lattices at certain places, as shown in Figure 24 (b). For example, the "roof" region of a printed part is typically the weakest, as the overhang angle $\theta$ there is very small. Figure 25 shows the distribution of the overhang angle $\theta$ of the bunny model, where the blue regions identify the "roofs" that are susceptible to material collapse and the density of the lattices in these places should be increased.

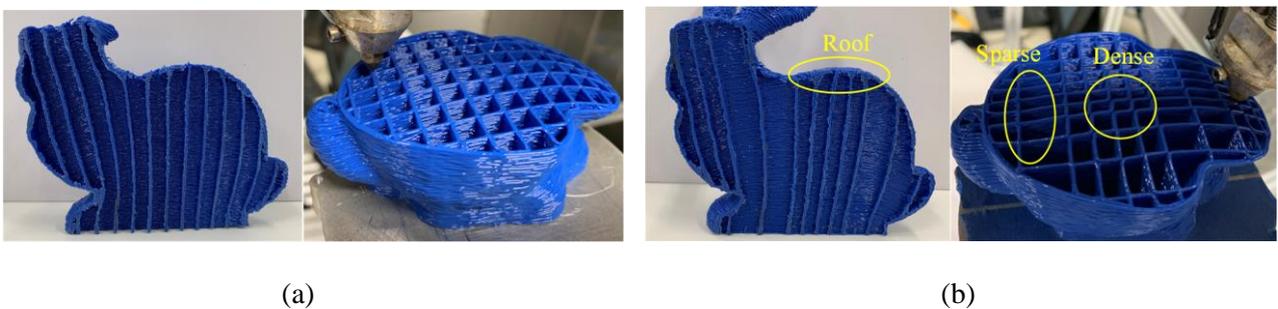

**Figure 24** Even and non-even lattices

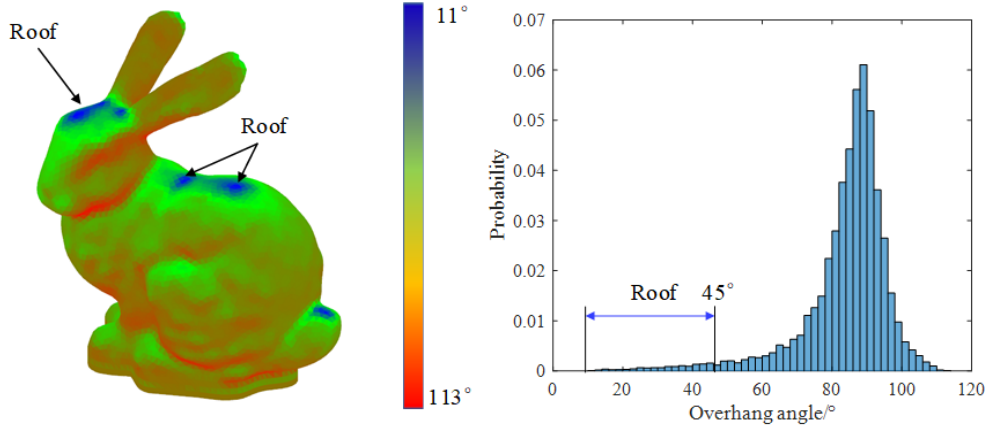

**Figure 25**  Distribution of the angle between the surface normal vector and the printing orientation

**4.2 Printing sequence optimization**

Next, we report the experimental results of three different printing sequence traversal algorithms, i.e., the benchmarking *LPT* and *DPT*, and our Algorithm 4 (A4), on a tree-structured part with three branches. As shown in Figure 26, the part is automatically decomposed into 151 infilling layers by our Algorithm 1 with the geodesic distance interval set at 0.6 mm, and the total number of the connected sub-graphs of the infilling layers is 311. Table 3 and Figure 27 show the simulation results of different cases. When the nozzle angle (which is denoted by NA that measures the size of the nozzle) is 75°, the printing sequence generated by the *LPT* is collision-free, which requires 162 retractions and the total air-move path length is 2382 mm. However, the *DPT* fails to generate a collision-free printing sequence, although the number of retractions (only 2) and the total air-move path length (only 101 mm) would be ideal. The proposed Algorithm 4 successfully plans a collision-free printing sequence with fewer retractions (24) and a shorter path length (654 mm) as compared with those of *LPT*. As expected, the number of nozzle retractions and the air-move path length are inversely related with the size of the nozzle (see Figure 27). Because the calculation of PCGs for each sub-graph involves collision check, the time complexity of A4 is the largest, i.e., $O(k^2*m)$, where $k$ and $m$ are the number of sub-graphs and the average number of vertices of each sub-graph, respectively. Figure 28 shows some snapshots of the actual printing processes of the A4 and *LPT* printing paths when the nozzle angle is 45°. (Note that we did not compare with the *DPT* path since it failed to print due to the unresolvable collisions.) The comparison results clearly confirm that, when compared to the *LPT* path, the filament drag problem is mitigated considerably by our A4 path owing to its significantly reduced nozzle retractions, thus leading to a much higher printing quality.

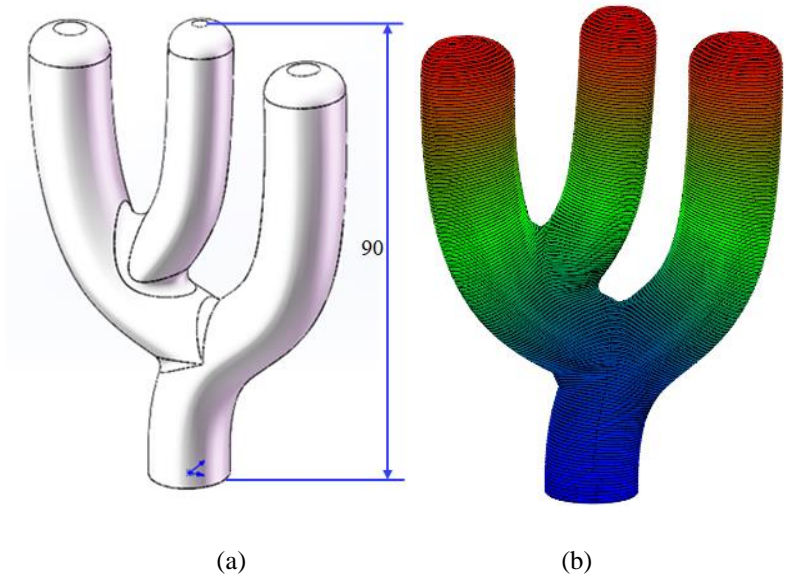

(a)                  (b)

**Figure 26** A three-branch model and its IGDSs: (a) model; (b) IGDSs.

Table 3 Simulation results of the three traversal algorithms

| Algorithm | Number of retractions | Air-move path length (mm) | Is collision-free or not | Running time (s) |
|---|---|---|---|---|
| LPT, NA: 75° | 162 | 2382 | Yes | 0 |
| DPT, NA: 75° | 2 | 101 | No | 0 |
| A4, NA: 75° | 24 | 654 | Yes | 271 |
| A4, NA: 60° | 17 | 371 | Yes | 177 |
| A4, NA: 45° | 7 | 227 | Yes | 277 |
| A4, NA: 30° | 5 | 146 | Yes | 154 |
| A4, NA: 15° | 4 | 125 | Yes | 155 |
| A4, NA: 1° | 3 | 108 | Yes | 242 |

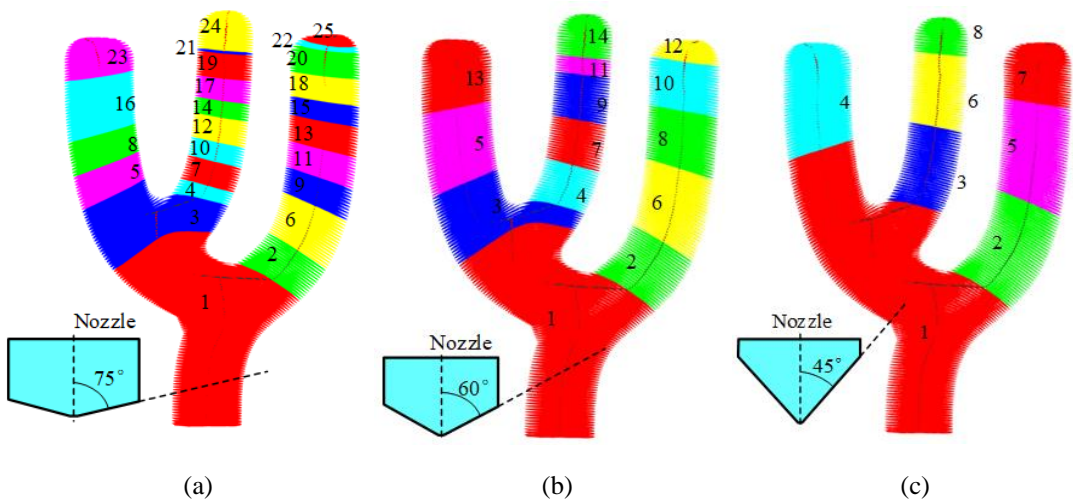

(a)                  (b)                  (c)

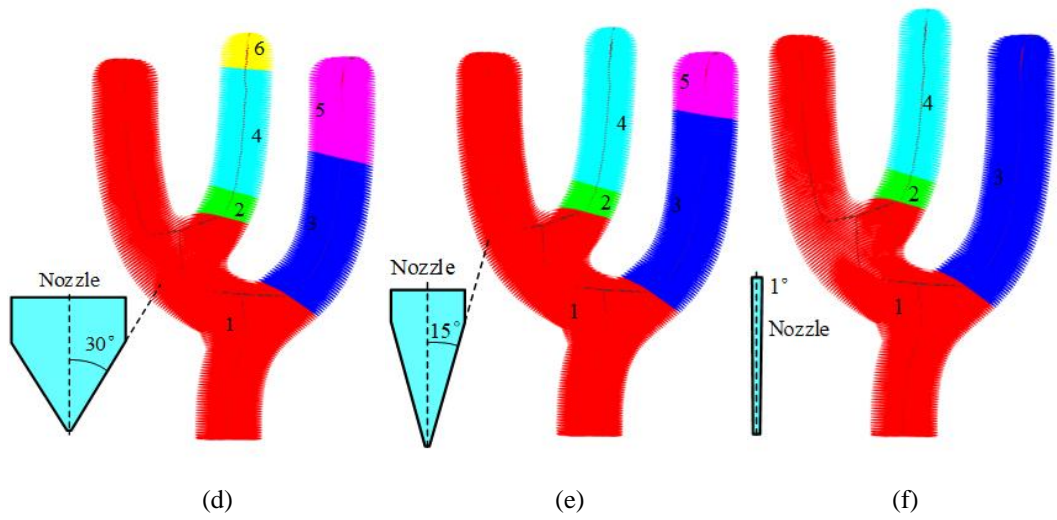

(d)            (e)            (f)

**Figure 27** Printing sequences generated by A4 under different nozzle angles: (a) 75°; (b) 60°; (c) 45°; (d) 30°; (e) 15°; (f) 1°.

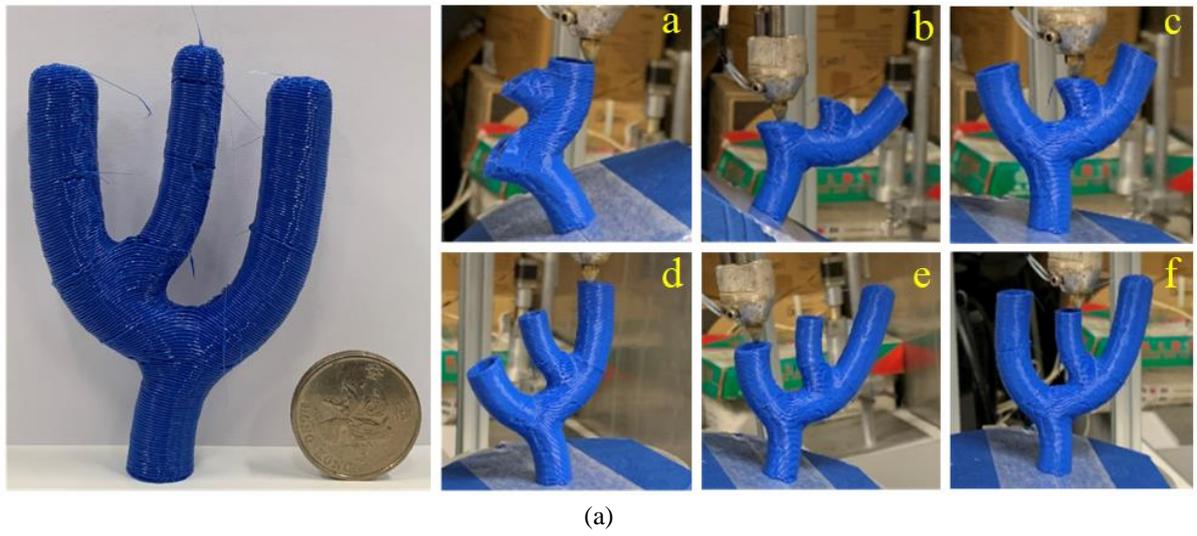

(a)

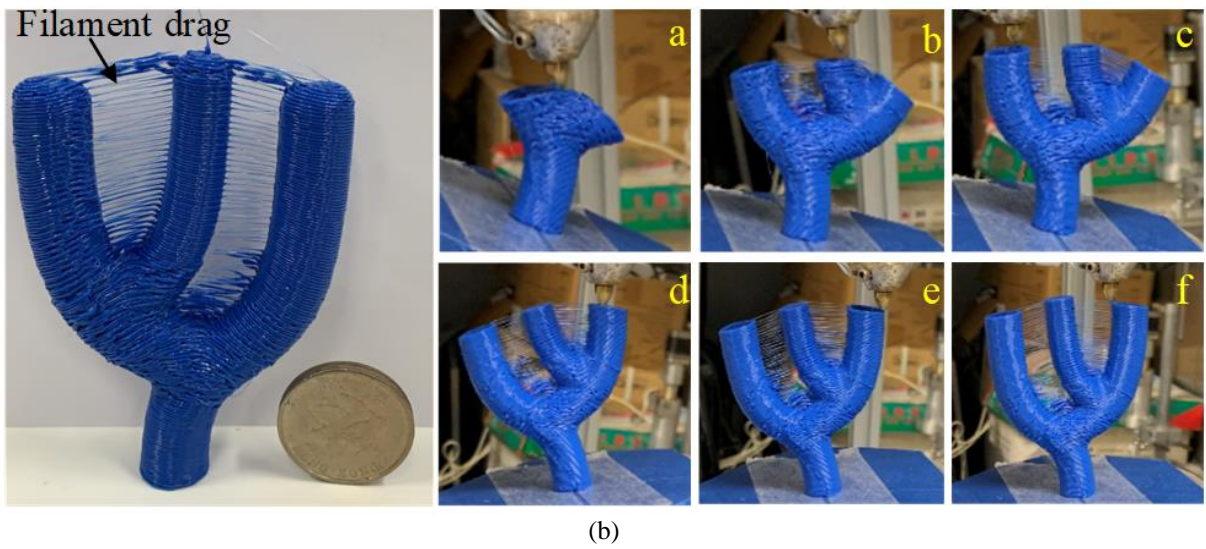

(b)

**Figure 28** Actual printing processes when the nozzle angle is 45°: (a) A4, printing path length is 39946 mm, printing time is150 min; (b) LPT, printing path length is 42101 mm, printing time is 155 min.

## 5. Conclusion

  This paper is motivated by the need of an automatic infill structure generation method for an arbitrary freeform solid part, that will ensure that both the generated infills and the given part boundary can be printed without any support under the continuous multi-axis printing configuration. Towards this objective, three mutually orthogonal geodesic distance fields embedded in the volume of the part are established. The iso-geodesic distance surfaces (IGDSs) of these three fields naturally form the curved layers of the part and the lattice infill structure, which, by aligning the nozzle orientation with the surface normal of the curve layers, guarantee that the overhang angles at both the part surface boundary and the infills are within the self-support range and thus eliminate the need of extra support. To avoid excessive nozzle retractions when printing the infills, the lattice infill pattern in each layer is first trimmed to an Eulerian graph and then a continuous printing path is constructed by using Fleury's algorithm. In addition, we present a printing sequence optimization algorithm for establishing a total ordering of the connected lattice infills which, while respecting the collision-free requirement, tries to minimize the air-move path length of the nozzle. The results of both computer simulation and physical printing experiments have given a positive confirmation of the proposed methods.

  Regarding the limitations and future research, the sub-graphs of the generated infills by our method cannot always maintain the convexity if the part has a complicated topology, which may cause local collisions when the nozzle angle is large. It is conceivable that, even under the most conservative LPT, there can be cases when the collision simply cannot be avoided on a skeleton tree. One solution to this problem is using a slender nozzle to reduce the potential of local collision. On the other hand, as there are many ways to decompose a solid and generate curved slicing layers (e.g., [17] and [18]), a solid that fails our 3D geodesics-based curved-layer slicing and printing sequencing algorithm might well be printable without collisions under other strategies of slicing and printing sequencing. One plausible solution is that, rather than given a base, we freely select a base (including both its location and the size) on the boundary of the solid to define the $\gamma$-geodesic distance field and the corresponding lattice infill patterns so that their corresponding skeleton tree will be printable at least under LPT. Alternatively, for a given solid we could combine the proposed 3D geodesics-based volume decomposition with other types of decomposition and come up with a different set of curved slicing layers and their printing sequence that are better in dealing with the collision. All these will be our future research topics.

**6. Acknowledgement**

This work is supported by Hong Kong RGC-GRF/16200819.